\documentclass[nofootinbib,preprintnumbers,amsmath,amssymb,aps,prc,twocolumn, showkeys,floatfix,superscriptaddress]{revtex4-1}

\usepackage{amsmath,graphics,color}
\usepackage{graphicx} 
\usepackage{graphics}
\usepackage{hyperref}
\usepackage{bm}
\usepackage{calligra}
\usepackage[T1]{fontenc}
\usepackage{amssymb}
\usepackage{amsfonts}
\usepackage{mathrsfs}  
\usepackage{dsfont}
\usepackage{footmisc}
\usepackage[normalem]{ulem}
\usepackage{xcolor}
\usepackage{caption}
\usepackage{subcaption}
\usepackage{ragged2e} 
\DeclareCaptionFormat{myformat}{\justifying#1#2#3}
\graphicspath{{figures/}}

\newcommand{\Tsw}{T_{\text{sw}}}
\newcommand{\trento}{\texttt{T$_\mathrm{R}$ENTo}}

\newcommand{\xv}{\mathbf x}
\newcommand{\SMASH}{\texttt{SMASH}}

\DeclareMathOperator{\sign}{sign}

\begin{document}



\title{The shape of transverse momentum spectra in hybrid hydrodynamic models}

\author{Thiago S. Domingues}
\email{thiago.siqueira.domingues@usp.br}
\affiliation{
 Instituto de F\'isica, Universidade de S\~ao Paulo, R. do Mat\~ao, 1371, S\~ao Paulo, Brazil, 05508-090
}%

\author{Fernando G. Gardim}
\affiliation{Instituto de Ci\^encia e Tecnologia, Universidade Federal de Alfenas, 37715-400 Po\c cos de Caldas, MG, Brazil
}

\author{Cicero D. Muncinelli}
\affiliation{Universidade Estadual de Campinas (Unicamp), R. S\'ergio Buarque de Holanda, 777, Campinas, Brazil, 13083-859
}

\author{Andre V. Giannini}
\affiliation{%
  Faculdade de Ciências Exatas e Tecnologia, Universidade Federal da Grande Dourados, 
  Dourados, MS, Brazil, 79804-970
}
\affiliation{Departamento de F\'isica, Universidade do Estado de Santa Catarina, 89219-710 Joinville, SC, Brazil}

\author{Gabriel S. Denicol}
\affiliation{%
  Instituto de F\'isica, Universidade Federal Fluminense,
  Av. Milton Tavares de Souza, Niter\'oi, Brazil, 24210-346,
}%

\author{Tiago Nunes da Silva}
\affiliation{%
 Departamento de F\'{i}sica, Centro de Ciências Físicas e Matemáticas, Universidade Federal de Santa Catarina, Campus Universit\'{a}rio Reitor Jo\~{a}o David Ferreira Lima, Florian\'{o}polis, Brazil, 88040-900}%

\author{David D. Chinellato}
\affiliation{Stefan Meyer Institute for Subatomic Physics of the Austrian Academy of Sciences, Wiesingerstraße 4
1010 Vienna, Austria}

\author{Giorgio Torrieri}
 \affiliation{%
  Universidade Estadual de Campinas (Unicamp), R. S\'ergio Buarque de Holanda, 777, Campinas, Brazil, 13083-859
}

\author{Mauricio Hippert}
 \affiliation{%
  Centro Brasileiro de Pesquisas F\'isicas, Rua Dr. Xavier Sigaud 150, Rio de Janeiro, RJ, 22290-180, Brazil
}

\author{Jun Takahashi}
 \affiliation{%
  Universidade Estadual de Campinas (Unicamp), R. S\'ergio Buarque de Holanda, 777, Campinas, Brazil, 13083-859
}

\author{Matthew Luzum}
\email{mluzum@usp.br}
\affiliation{
 Instituto de F\'isica, Universidade de S\~ao Paulo, R. do Mat\~ao, 1371, S\~ao Paulo, Brazil, 05508-090
}%

\collaboration{The ExTrEMe Collaboration}

\date{\today}
\begin{abstract}
We study the scaled transverse momentum spectra over a wide parameter space of state-of-the-art hydrodynamic simulation models in order to learn what information can be obtained from the shape of identified-particle spectra --- previously observed to be surprisingly universal across centrality and collision systems in both experimental data and hydrodynamic simulations.  We study its sensitivity to each of 17 model parameters in the context of 4 different models for particlization when switching from the hydro description to the kinetic theory afterburner. We find that the strongest sensitivity is to parameters relating to bulk viscosity, free-streaming time, and the \trento{} nucleon width parameter $w$.
However, we find that the model generally has surprisingly little flexibility in describing the scaled spectrum observable, despite the large number of parameters.

Within this small range of parameter dependence, we further find significant tension in a simultaneous description of momentum-integrated observables.  In particular, while the mean transverse momentum prefers a large value of the nucleon width parameter $w$, a small value is required to obtain scaled spectra that are consistent with experimental measurements.
We speculate on the origin of these model tensions and possible missing physics in the commonly-used \trento{}+free streaming+hydro+afterburner simulation model.
\end{abstract}
\maketitle
\section{Introduction}
\label{sec:intro}
Ultrarelativistic heavy-ion collisions at facilities such as the LHC create extreme conditions of temperature and energy density, producing a deconfined state of nuclear matter known as the Quark–Gluon Plasma (QGP)~\cite{Heinz:2013th,Busza:2018rrf}. A key achievement in the field has been characterizing the collective behavior of the QGP with relativistic viscous hydrodynamics, which accurately describes a broad range of observables, including anisotropic flow coefficients \(v_n\) and identified particle spectra~\cite{Heinz:2013th,Niemi:2015qia}. These measurements encode detailed information on the equation of state, transport coefficients such as shear and bulk viscosity, and the effects of initial-state fluctuations~\cite{Bernhard:2016tnd}.

A fundamental observable in heavy-ion collisions is the 
transverse momentum spectrum of produced particles. This $p_T$-differential spectrum encodes information about
particle production and the collective expansion of the medium, though it is commonly summarized through its 
lowest moments---the charged particle multiplicity $N_{\rm ch}$ and the average transverse momentum $\langle p_T\rangle$~\cite{Novak:2013bqa}.

While these $p_T$-integrated quantities have been extensively used in Bayesian analyses to constrain medium properties~\cite{Bernhard:2016tnd,Nijs:2021clz, JETSCAPE:2020mzn, JETSCAPE:2020shq}, the full shape of the \(p_T\) spectra contains complementary, highly nontrivial information that is sensitive to late-stage hadronic interactions, the nature of the QGP–hadron gas transition, and semi-hard processes~\cite{Nijs:2020ors, Nijs:2020roc}. Specifically, the  calibration made by~\cite{JETSCAPE:2020mzn} adopted a deliberately conservative strategy by restricting the calibration to low-$p_T$ observables, specifically $p_T$-integrated quantities. This choice reduces the influence of model-dependent uncertainties that become increasingly significant at higher transverse momentum. This approach is supported by evidence that the bulk of the physical information contained in the low-$p_T$ spectra (approximately $p_T \lesssim 1.5\,\mathrm{GeV}$) is already encoded in $p_T$-integrated observables such as the multiplicity and mean transverse momentum~\cite{Novak:2013bqa}. In contrast, the higher-$p_T$ region ($p_T \gtrsim 1.5\,\mathrm{GeV}$) introduces substantially larger theoretical uncertainties, particularly due to the sensitivity to the viscous correction functions applied at particlization~\cite{JETSCAPE:2020mzn, JETSCAPE:2020shq}.

To systematically isolate and study the $p_T$-spectra shape independent of global quantities, it was introduced the dimensionless scaled spectrum~\cite{Muncinelli:2024izj}
\begin{equation}
  U(x_T) \equiv \frac{\langle p_T \rangle}{N} \frac{dN}{dp_T} = \frac{1}{N} \frac{dN}{dx_T}\,,
  \quad x_T = \frac{p_T}{\langle p_T \rangle} \,,
  \label{eq:Udef}
\end{equation}
which removes the multiplicity scaling and rescales transverse momenta by the mean momentum, \(\langle p_T \rangle\). This construction reveals an apparent universal shape of \(U(x_T)\) across different collision systems (Pb–Pb, Xe–Xe, p–Pb) and centralities, persisting up to \(x_T \sim 3\), with systematic deviations appearing only at higher \(x_T\), where semi-hard processes dominate, and in very peripheral events. 
In hydrodynamic simulations, this universality arises from an \emph{event-by-event} universal shape ~\cite{Muncinelli:2024izj}.%
\footnote{We note that other works have already mentioned a universal scaling in different contexts~\cite{Rybczynski:2012vj, Andres:2012ma, Khachatryan:2022qrc, Moriggi:2020zbv, Andres:2016mla, Andres:2014xka, Schaffner-Bielich:2001dlb}.}

Despite its potential as a novel hydrodynamic signature, exploring the constraints from \(U(x_T)\) into state-of-the-art Bayesian model calibrations remains largely unexplored. 
In this work, we address this gap by extending previous Bayesian analyses from the JETSCAPE Collaboration \cite{JETSCAPE:2020shq, JETSCAPE:2020mzn} --- which couple \trento{} initial conditions, free-streaming pre-equilibrium dynamics, MUSIC viscous hydrodynamics, iS3D particlization, and SMASH hadronic transport ---  
to include \(U(x_T)\) alongside multiplicity, \(\langle p_T \rangle\), and anisotropic flow data from the ALICE Collaboration.
Our study aims to:
\begin{enumerate}
\item Determine which aspects of hydrodynamic models affect the scaled spectra via a global sensitivity analysis over a large parameter space, as well as different models for particlization.
\item Explore the ability of the hybrid model to describe experimental data for the scaled spectra, both individually and simultaneously with $p_T$-integrated observables via Bayesian inference.
\end{enumerate}
By investigating this newly characterized scaled spectra in global model calibration, we first demonstrate that this novel observable displays a weak sensitivity to the model parameters. Furthermore, we show that current state-of-the-art models struggle to reproduce this observable, indicating that it probes physics not well described in hybrid simulations. Thus, our work represents an important step toward a more detailed quantitative understanding of collective dynamics and particle production in ultrarelativistic heavy-ion collisions.

The paper is structured as follows. In Sec.~\ref{sec:hybrid_simulation}, we describe the JETSCAPE framework to simulate the soft sector, including four different approaches to viscous corrections in particlization, which we systematically investigate in this work. In Sec.~\ref{sec:analysis_setup}, we present the methods used for our global analysis using the scaled spectra. Our surrogate models and their training are described in  Sec.~\ref{subsec:gp_emulator_pcs}, while the details of our model-to-data calibration are explained in Sec.~\ref{subsec:bayesian_framework}. 
In Sec.~\ref{sec:results}, we present the results of our analysis. 
We start by displaying predictions for Pb--Pb collisions at 2.76 TeV within the 14-moment Grad approach to viscous corrections, compared to experimental data from the ALICE Collaboration in Sec.~\ref{subsec:model_vs_data}, which is followed by a global sensitivity analysis across all the model parameters in Sec.~\ref{subsec:sensitivity_analysis_params}.  
In Sec.~\ref{subsec:viscous_corrections}, we discuss the sensitivity of the scaled spectra to the modeling of out-of-equilibrium corrections $\delta f$ to the momentum distribution of particles, comparing the four different approaches we investigate. 
Finally, we summarize and discuss our conclusions in Sec.~\ref{sec:conclusions}. In the Appendix, we present details about the PCA decomposition, the training and validation of our emulator, and our global sensitivity analysis over different models for the out-of-equilibrium  $\delta f$ corrections.

\section{Hybrid Simulation Framework}
\label{sec:hybrid_simulation}
To access the scaled spectra, we use theoretical calculations performed by the JETSCAPE Collaboration in earlier works~\cite{JETSCAPE:2020mzn, JETSCAPE:2020shq, Everett:2021ruv}. These calculations were designed to study the soft sector of heavy-ion collisions for different collision systems and energies: Pb--Pb at 2.76 TeV, Au--Au at 0.2 TeV, and Xe--Xe at 5.44 TeV~\cite{Everett:2021ruv}. These works demonstrated robust agreement with $p_T$-integrated experimental data across these systems and energies.

In this section, we provide a concise overview of each JETSCAPE hybrid model component, with further details available in~\cite{JETSCAPE:2020shq}. Given a point in the high-dimensional parameter space $\boldsymbol{\theta}$, this hybrid multi-stage framework consists of five stages, which will be discussed in more detail in the following subsections.
\subsection{Initial energy deposition}
\label{subsec:initial_deposition}
The initial energy deposition profile at proper time $\tau = 0^+$, immediately after the nuclear collision, is produced using the \trento\ parametric model~\cite{Moreland:2014oya}, and is then input into  a free-streaming expansion phase~\cite{Liu:2015nwa, Broniowski:2008qk}.  The \trento\ model creates the initial energy density distribution in the transverse plane based on the positions of participating nucleons, sampled from the nuclear density distributions of the colliding nuclei. This parameterization of the energy density in the transverse plane is given in terms of a reduced thickness function, $T_R$: 
\begin{eqnarray}
    \bar\epsilon(\mathbf{x}_\perp) 
    &\equiv& \frac{dE}{d\eta_s d^2\mathbf{x}_\perp} 
    = N T_R(\mathbf{x}_\perp; p),  
\label{eq:TrentoEd}
\\
\label{eq:TR}
    T_R(\mathbf{x}_\perp; p) &=& \left(\frac{T^p_A(\mathbf{x}_\perp) + T^p_B(\mathbf{x}_\perp)}{2}\right)^{1/p},
\end{eqnarray}
where $N$ is an overall normalization factor that sets the absolute scale of the initial energy deposition, directly affecting the overall multiplicity of the produced particles. 

The thickness functions $T_A$ and $T_B$ represent the participant nucleon densities of nuclei A and B, respectively, while the reduced thickness function $T_R$ of all participants defines the initial energy density. 
The mapping from $T_A$ and $T_B$ to $T_R$, and ultimately to the initial energy deposition $\bar\epsilon(\mathbf{x}_\perp)\equiv\lim_{\tau\to0^+}\tau\epsilon(\tau,\mathbf{x}_\perp,\eta_s{=}0)$, is controlled by the free parameter $p$. It determines how the thickness functions of the two colliding nuclei are combined: $p = 1$ corresponds to the Wounded Nucleon model (arithmetic mean), $p = 0$ gives the geometric mean (similar to the KLN model), and negative values approach the EKRT model. The value of $p$ effectively controls the relative contribution of each nucleus to energy production. Since $p$ is a mapping between different ways to combine the participant thickness functions, we can call it the generalized mean (see Table~\ref{tb:prior_table}).

The nucleon's effective size in $T_A$ and $T_B$ is controlled by the effective nucleon width, \(w\), measured in fm, which determines the initial state granularity~\cite{Moreland:2019szz}. The nucleon thickness function is modeled by a three-dimensional Gaussian function, with $w$ being the width of the distribution:
\begin{eqnarray}
    \rho(\mathbf{x}_\perp) &=& \int_{-\infty}^{\infty} \frac{dz}{\left(2\pi w^2\right)^{3/2}}\exp\left(-\frac{\mathbf{x}_\perp^2+z^2}{2w^2}\right),
\label{eq:nucleon_width}    
\end{eqnarray}
The nucleon width $w$ also controls how nucleon-nucleon interactions are resolved.

Nucleon position correlations are present through a free parameter $d^3_{\rm min}$, which models the minimum distance (in fm) to be held between nucleons in the same nucleus, setting an exclusion volume~\cite{ALICE:2013hur, Loizides:2014vua, Rybczynski:2013yba}:
\begin{equation}
  \label{eq:min_dist}
  |\xv_i - \xv_j| > d_\text{min}.
\end{equation}
Finally, extra fluctuations are controlled by the parameter $\sigma_k$, which specifies the standard deviation of a  $\Gamma$-distribution with a unit mean to model fluctuations in the nucleon and its energy deposition.

For the present study, we used the generated 2,500 minimum bias initial condition events for Pb--Pb collisions at 2.76 TeV from~\cite{JETSCAPE:2020shq}. Each initial condition represents a distinct collision event with its own fluctuating geometry, determined by the random sampling of nucleon positions within the colliding nuclei. The prior distribution associated with each parameter is displayed in Table~\ref{tb:prior_table}.
\subsection{Pre-equilibrium}
\label{subsec:pre-eq}
The initial energy density generated in \trento\ is further evolved through a simple free-streaming stage to simulate the pre-equilibrium stage of the collision. Free-streaming propagates the initial phase-space distribution by solving the collisionless Boltzmann equation for massless particles \cite{deGroot:80relativistic}.
\begin{equation}
        P^{\mu} \partial_{\mu} f(X;P) = 0,
\label{eq:Boltzmann_free_streaming}   
\end{equation}
with $P^\mu=(|\boldsymbol{p}|,\boldsymbol{p})$ being the 4-momentum and $X^\mu=(t,\boldsymbol{x})$ the space-time coordinate. 

%
%
%
%
%
%
This evolution only stops at a longitudinal proper time \(\tau_{\rm fs}\) when the energy-momentum tensor is matched to the viscous hydrodynamics through conditions called Landau matching~\cite{LandauLifshitzFluids}. The time to initialize hydrodynamics, or the free-streaming time, is  parameterized by a dependence on the initially deposited transverse energy density $\bar\epsilon$ (Eq.~\ref{eq:TrentoEd}): 
\begin{equation}
   \tau_{\rm fs} = \tau_R \left(\frac{\langle \bar\epsilon \rangle}{\bar\epsilon_R}\right)^{\alpha},
\label{eq:taufs}
\end{equation}
where $\tau_R$ is an overall factor for the free-streaming model duration, and the free parameter $\alpha$ controls the dependence of the free-streaming time on the average energy density over the transverse plane, defined as 
\begin{equation}
\label{eq:eps_ave}
    \langle \bar\epsilon \rangle \equiv \frac{\int d^2x_\perp\, {\bar\epsilon}^2(\mathbf{x}_\perp)}{\int d^2x_\perp\, \bar\epsilon(\mathbf{x}_\perp)}.
\end{equation}
Following~\cite{JETSCAPE:2020shq}, we set $\bar\epsilon_R = 4.0$\,GeV/fm$^2$ as a reference scale. 
\subsection{Hydrodynamic model}
\label{subsec:hydro_model}
After the pre-equilibrium stage, the energy-momentum tensor is evolved using second-order relativistic viscous hydrodynamics~\cite{Schenke:2010nt, Schenke:2010rr, Paquet:2015lta, Denicol:2012es, Shen:2014vra}, employing a dissipative fluid dynamics code called MUSIC~\cite{Schenke:2010nt,Shen:2014vra}. Our analysis focuses on the midrapidity region, which can be approximated using a longitudinal boost-invariant expansion. 

The equation of state employed in this work is from the HotQCD Collaboration~\cite{HotQCD:2014kol},
which combines lattice predictions at high temperatures with a hadron resonance gas description at lower temperatures~\cite{Bernhard:2018hnz}.

Furthermore, we use the following temperature-dependent parameterizations for the shear and bulk viscosities, $\eta$ and $\zeta$:
\begin{equation}
\label{eq:shear_positivity}
    \frac{\eta}{s}(T) = \max\left[\left.\frac{\eta}{s}\right\vert_{\rm lin}\!\!\!(T),0\right],
\end{equation}
with
\begin{eqnarray}
    \left.\frac{\eta}{s}\right\vert_{\rm lin}\!\!\!(T) &=& a_{\rm low}\, (T{-}T_{\eta})\, \Theta(T_{\eta}{-}T)+ (\eta/s)_{\rm kink}
\nonumber\\
    && +\, a_{\rm high}\, (T{-}T_{\eta})\, \Theta(T{-}T_{\eta}),
\label{eq:shear_viscosity_param}    
\end{eqnarray}  
and a skewed Cauchy distribution,
\begin{eqnarray}
    \frac{\zeta}{s}(T) &=& \frac{(\zeta/s )_{\max}\Lambda^2}{\Lambda^2+ \left( T-T_\zeta\right)^2},\\
    \Lambda &=& w_{\zeta} \left[1 + \lambda_{\zeta} \sign \left(T{-}T_\zeta\right) \right]\nonumber.
\label{eq:bulk_viscosity_param}    
\end{eqnarray}
Finally, another free parameter of the model is the shear relaxation time, which can be parameterized by the following temperature dependence.
\begin{equation}
    T \tau_\pi(T)= b_{\pi}\frac{\eta}{s}(T)
\label{eq:shear_relax_time}    
\end{equation}
%
%
%
\subsection{Particlization}
\label{subsec:particlization}
After the hydrodynamic stage, the system is described using relativistic transport theory as a distribution of particles whose dynamics are described via local collisions and decay probabilities. In order to initialize the transport simulation, we first need to convert the fluid cells into particles with positions and momenta; this transition is performed on a hypersurface at a constant temperature $\Tsw$. In the JETSCAPE framework, the Cooper–Frye prescription \cite{PhysRevD.10.186} is implemented with iS3D to convert hydrodynamic fluid cells to particles at an isothermal switching hypersurface according to a specific particlization model.

Since this transition from hydrodynamics to hadronic degrees of freedom is not a well-defined problem, the macroscopic description cannot provide any additional information about the microscopic dynamics at the switching surface. While the local-equilibrium distribution is well defined, we cannot determine exactly the nonequilibrium correction to the momentum distribution of particles, \(\delta f\), at the switching hypersurface.
The shape of \(\delta f\) in the Cooper–Frye prescription encodes how viscous stresses modify the local particle distributions, but its determination is ambiguous.
We investigate this ambiguity using 4 different models to determine the \(\delta f\) of the hadronic phase-space distributions in terms of the $T^{\mu \nu}$ components. We  briefly summarize these four models  below. A more detailed description can be found in Ref.~\cite{McNelis:2019auj}.

%
\subsubsection{Linear viscous corrections: 14-moments (Grad) and Chapman-Enskog (CE)}
\label{subsubsec:linear_viscous_corrections}
The relativistic 14-moments method (Grad's method)~\cite{Grad:1949zza, Dusling:2009df, McNelis:2019auj, Teaney:2003kp, Monnai:2009ad, Dusling:2011fd, Denicol:2012cn} and the linearized Chapman-Enskog expansion in the relaxation time approximation (CE)~\cite{Anderson:1974nyl, Jaiswal:2014isa} are linearized corrections to the dissipative currents $\pi^{\mu \nu}$ and $\Pi$. We expect these approximations to be valid only when the deviations from the thermal equilibrium distribution are small.
\begin{description}
  \item[The 14-moment Grad]  
  We expand the correction to the local equilibrium distribution function in powers of hadronic momentum. 
  \begin{equation}
    \delta f_i =  f_{\text{eq}, i}  \bar{f}_{\text{eq}, i} c_{\mu\nu}P^{\mu}P^{\nu}
  \label{eq:viscous_correction_equation}  
  \end{equation}
  Here, $f_{\text{eq}, i}$ is the corresponding local-equilibrium distribution function of the $i$--th hadronic species and $\bar{f}_{\text{eq}, i} \equiv 1{-}\Theta f_{\text{eq}, i}$, where $\Theta$ is 1 for fermions and ${-}1$ for bosons. The species-independent coefficients $c_{\mu\nu}$ are related to the viscous stresses via the Landau matching conditions to yield the expression for the Grad viscous correction:
  \begin{eqnarray}
    \delta f^{\text{Grad}}_i = f_{\text{eq}, i} \bar{f}_{\text{eq}, i} \bigl[ \Pi \left( A_{T}m_i^2{+}A_E(u{\cdot}P)^2  \right)\ && 
    \nonumber\\
    + A_{\pi}\pi^{\mu\nu}P_{\langle \mu}P_{\nu \rangle}  \bigr].
    \label{eq:Grad_equation}
  \end{eqnarray}
  The terms $A_{T}$, $A_{E}$, and $A_{\pi}$ are thermodynamic moments of the equilibrium function~\cite{McNelis:2019auj}; $m_i$ is the hadronic mass of each species $i$, 
  $P_{\langle \mu}P_{\nu \rangle}=\Delta_{\mu\nu}^{\alpha\beta}\, P_\alpha P_\beta$, and $\Delta_{\mu\nu}^{\alpha\beta}$ is the transverse-traceless projector. 
    
  \item[Chapman-Enskog RTA]  
  We derive \(\delta f\) by iteratively solving the Boltzmann equation in the relaxation‐time approximation, giving a momentum‐linear correction:
  \begin{eqnarray}
    \delta f^{\text{CE RTA}}_i = f_{\text{eq}, i} \bar{f}_{\text{eq}, i}
    \left[  \frac{\Pi}{\beta_{\Pi}}\left( \frac{(u{\,\cdot\,}P) \mathcal{F}}{T^2} - \frac{P{\,\cdot\,}\Delta{\,\cdot\,}P}
    {3(u{\,\cdot\,}P)T}\right) \right.\ && 
    \nonumber\\ 
    \left. + \frac{ \pi_{\mu\nu}P^{\langle \mu}P^{\nu \rangle}}{2\beta_{\pi}(u \cdot P)T} \right].&&\quad
    \label{eq:CE_equation}
  \end{eqnarray}
  The definitions for $\mathcal{F}$, $\beta_\pi$, and $\beta_\Pi$ can be found in~\cite{McNelis:2019auj}. 
\end{description}
\subsubsection{Exponentiated viscous corrections: Pratt-Torrieri-McNelis (PTM) and Pratt-Torrieri-Bernhard (PTB)}
\label{subsec:PTM_PTB_models}
\begin{description}
    \item[Pratt–Torrieri–McNellis (PTM)]
    We can define the Pratt-Torrieri-McNellis (PTM)~\cite{Pratt:2010jt, McNelis:2019auj} distribution as follows:
    \begin{equation}
    f_{\text{PTM}} = \mathcal{Z} \left[ \exp\left(\frac{ \sqrt{|\mathbf{P'}|^2 + m^2} }{T + \beta_{\Pi}^{-1} \Pi \mathcal{F}}\right) + \Theta \right]^{-1}.
    \label{eq:PTM_equation}
    \end{equation}
    where $\mathcal{Z}$ is a scaling factor that is $\Pi$ and mass dependent on hadrons. 
    \item[Pratt–Torrieri–Bernhard (PTB)]
    Finally, we can define the Pratt-Torrieri-Bernhard viscous correction~\cite{Bernhard:2016tnd,Pratt:2010jt} as follows:
    \begin{equation}
    f_{\text{PTB}} = \frac{\mathcal{Z}_{\Pi}}{\text{det} \Lambda} \left[\exp\left(\frac{ \sqrt{|\mathbf{P'}|^2 + m^2} }{T}\right) + \Theta \right ]^{-1},
    \label{eq:PTB_equation}
    \end{equation}
    where here, $\mathcal{Z}_{\Pi}$ is another scaling factor~\cite{Bernhard:2018hnz, McNelis:2019auj} that is $\Pi$ dependent and species-independent.  
\end{description}
\subsection{Hadronic Transport}
\label{subsec:hadronic_transport}
Once the fluid is converted to particles, the emitted hadrons can scatter, form resonances, and decay. This process was simulated by the JETSCAPE Collaboration using the kinetic evolution code \SMASH{}~\cite{SMASH:2016zqf, smash_code, Sciarra:2024gcz, Elfner:2025ojd}, which solves coupled Boltzmann equations for hadronic species $i$ and a collision term $C[f_i]$ describing all scatterings, resonances, and decays:
\begin{equation}
    P^{\mu} \partial_{\mu} f_i(x;P) = C[f_i].
\label{eq:Boltzmann_equation}    
\end{equation}
This multi-stage hybrid model defines centrality classes using $dN_{\rm ch}/d\eta$ to order the minimum bias events. Here, we calculate the scaled spectra in the same centrality and $x_T$ ($p_T/\langle p_T \rangle$) bins used by the experimental collaborations.
Including these four \(\delta f\) prescriptions in our Bayesian framework allows us to quantify the systematic uncertainty associated with the particlization step and assess how different viscous corrections impact the reduced spectra. 

This concludes a brief overview of the JETSCAPE framework simulations used throughout this work. The statistical analysis framework employed to calibrate the universal spectra to experimental data is described next.

%
\section{Analysis Setup}
\label{sec:analysis_setup}
\subsection{Gaussian‐Process Emulation of Principal Components}
\label{subsec:gp_emulator_pcs}
Following previous works~\cite{Bernhard:2016tnd, JETSCAPE:2020shq}, Gaussian processes (GPs) are an essential tool for Bayesian parameter estimation in heavy-ion collisions. They can efficiently interpolate between full model calculations at a few design points in a high-dimensional parameter space, providing an estimation of their own uncertainty. 
To efficiently emulate the full-scaled spectrum observable \(U(x_T)\), which for Pb--Pb at 2.76 TeV spans 41 normalized momentum bins across 7 centrality classes (total \(41 \times 7 = 287\) data points), we employ a two‐step procedure:
\paragraph{Dimensionality Reduction via PCA:}
we collect the scaled spectra from our \(N_{\rm dp}\) design points hybrid simulations into a matrix. 
\begin{equation}
\mathbf{X}\in\mathbb{R}^{N_{\rm dp}\times m},
\label{eq:design_matrix}    
\end{equation}
where each row $N_{\rm dp}$ corresponds to one design point in the parameter space (characterized by a parameter vector \(\boldsymbol{\theta}_i\)) and each column m corresponds to a particular \((x_T,\,{\rm cent})\) bin. After standardizing each column, we perform an eigen‐decomposition of the empirical covariance matrix:
\begin{equation}
\mathbf{C}=\frac{1}{N_{\rm dp}-1}\,\mathbf{X}^\mathsf{T}\mathbf{X}
\;\longrightarrow\;
\{\lambda_\alpha,\;\mathbf{e}_\alpha\}_{\alpha=1}^{m}.
\label{eq:covariance_decomposition}    
\end{equation}
We find that the first \(K=6\) principal components capture over 98\% of the total variance (see Appendix ~\ref{appendix:emulator_PCA}):
\begin{equation}
\frac{\sum_{\alpha=1}^{K=6} \lambda_\alpha}{\sum_{\alpha=1}^{m=287} \lambda_\alpha}\;>\;0.98.
\label{eq:principal_components}
\end{equation}
We thus project each simulation onto these six leading eigenvectors to obtain the principal component scores \(z_\alpha\), which represent the coordinates of each design point in the reduced-dimensional space:
\begin{equation}
z_\alpha(\boldsymbol{\theta}_i)
=\bigl[\mathbf{X}_i - \bar{\mathbf{X}}\bigr]\cdot\mathbf{e}_\alpha,
\quad \alpha=1,\dots,6,
\label{eq:pc_equation}    
\end{equation}
where \(\boldsymbol{\theta}_i\) denotes the model parameter vector at the \(i\)-th design point, \(\mathbf{X}_i\) is the corresponding spectrum (the \(i\)-th row of \(\mathbf{X}\)), and \(\bar{\mathbf{X}}\) is the mean spectrum over all design points.
\paragraph{Independent Gaussian‐Process Emulators:}
for each principal component coefficient \(z_\alpha\), we build an independent Gaussian‐process (GP) emulator:
\begin{equation}
z_\alpha(\boldsymbol{\theta})\;\sim\;\mathcal{GP}\bigl(\nu_\alpha(\boldsymbol{\theta}),\,k_\alpha(\boldsymbol{\theta},\boldsymbol{\theta}')\bigr),
\label{eq:pc_coefficient}    
\end{equation}
with:
\begin{itemize}
  \item \textbf{Mean function} \(\nu_\alpha(\boldsymbol{\theta}) = \mu_\alpha\), treated as a constant and set to the empirical average of \(z_\alpha\).
  \item \textbf{Covariance kernel}
  \begin{equation}
  k_\alpha(\boldsymbol{\theta},\boldsymbol{\theta}')
    = \sigma_\alpha^2
      \exp\Bigl[-\tfrac12\sum_{d=1}^n
      \bigl(\tfrac{\theta_d-\theta'_d}{\ell_{\alpha,d}}\bigr)^2\Bigr]
      + \delta_{\boldsymbol{\theta},\boldsymbol{\theta}'}\,\tau_\alpha^2,
  \label{eq:covariance_kernel}    
  \end{equation}
\end{itemize}
where \(n\) is the number of model parameters (17 in our case, see Table~\ref{tb:prior_table}), \(\{\ell_{\alpha,d}\}\) is the length scales, \(\sigma_\alpha^2\) is the signal variance, and \(\tau_\alpha^2\) is a nugget term accounting for residual noise. The hyperparameters \(\{\mu_\alpha,\sigma_\alpha,\ell_{\alpha,d},\tau_\alpha\}\) are determined by maximizing the marginal log‐likelihood of the training data \(\{(\boldsymbol{\theta}_i,\,z_\alpha(\boldsymbol{\theta}_i))\}\) using gradient‐based optimization~\cite{Bernhard:2018hnz, JETSCAPE:2020shq}.\footnote{Further information can be found in the Python library \href{https://scikit-learn.org/stable/modules/gaussian_process.html}{Scikit-learn documentation}}
\subsection{Bayesian Inference Framework}
\label{subsec:bayesian_framework}
In this section, we provide a brief overview of Bayesian inference in the context of relativistic heavy-ion collisions. Bayesian parameter estimation is a modern statistical framework for inferring unknown model parameters from data while coherently quantifying uncertainty. In heavy-ion collisions, it is now a standard approach to constrain the properties of the QGP~\cite{Petersen:2010zt, Novak:2013bqa, Sangaline:2015isa, Bernhard:2015hxa, Bernhard:2016tnd, Bernhard:2019bmu, Domingues:2024pom, Domingues:2025tkp} because it (i) naturally accounts for multiple sources of uncertainty, (ii) yields full posterior probability distributions for parameters, and (iii) provides principled propagation of uncertainty to predictions and derived quantities. 

Following the previous section, let $\boldsymbol{\theta} \in\Theta\subset\mathbb{R}^n$ denote the vector of model parameters in an n-dimensional parameter space, and let \(D\) denote all sources of data, including experimental (observables) and simulation data. Bayes' theorem updates prior beliefs about $\boldsymbol{\theta}$ using the likelihood of the data:
\begin{equation}
p(\boldsymbol{\theta}\mid D) \;=\; \frac{p(D\mid\boldsymbol{\theta})\,p(\boldsymbol{\theta})}{p(D)} \;=\; \frac{ \mathcal{L}(\boldsymbol{\theta})\,p(\boldsymbol{\theta}) }{ \displaystyle \int_\Theta \mathcal{L}(\boldsymbol{\theta})\,p(\boldsymbol{\theta})\,d\boldsymbol{\theta} }.
\label{eq:bayes}
\end{equation}
Here, \(p(\boldsymbol{\theta})\) is the \emph{prior} density encoding prior knowledge (or ignorance) about parameters, \(\mathcal{L}(\boldsymbol{\theta})=p(D\mid\boldsymbol{\theta})\) is the \emph{likelihood}, the probability of observing the data under the model with parameters \(\boldsymbol{\theta}\). Finally, \(p(D)\) is the \emph{evidence} (marginal likelihood), a normalizing constant that is important for model comparison but not required for parameter inference. As is common in heavy-ion collisions, we choose uniform distributions for the prior. 

For a model with $n$ parameters, their posterior \(p(\boldsymbol{\theta}\mid D)\) is an \(n\)-dimensional distribution describing plausible parameter configurations given the data and prior knowledge. The posterior is rarely analytically tractable for realistic simulators, so we \emph{sample} it numerically (e.g., with MCMC or nested sampling). Posterior samples can be used to determine credible intervals, and posterior predictive distributions can communicate our model-to-data constraints.

A careful likelihood must include all relevant sources of uncertainty:
\begin{equation}
D = \mathcal{M}(\boldsymbol{\theta}) + \delta_{\text{model}} + \varepsilon_{\text{exp}},
\label{eq:bayes_data}
\end{equation}
where \(\mathcal{M}(\boldsymbol{\theta})\) denotes the simulator output (deterministic or stochastic), \(\delta_{\text{model}}\) is model discrepancy (structural error), and \(\varepsilon_{\text{exp}}\) is experimental noise. A typical multi-variate Gaussian likelihood (after suitable transforms/standardization) is:
\begin{align}
\mathcal{L}(\boldsymbol{\theta})
\propto 
\frac{1}{\sqrt{(2\pi)^m\det\!\big(\,\Sigma_{\text{tot}}(\boldsymbol{\theta})\big)}}
\exp\!\left[
-\tfrac12\,\Delta^\top 
\Sigma_{\text{tot}}(\boldsymbol{\theta})^{-1}
\Delta
\right],
\label{eq:likelihood_function}
\end{align}
where  $\Delta = D - \mu_{\mathcal{M}}(\boldsymbol{\theta}),
\Sigma_{\text{tot}}(\boldsymbol{\theta})
= \Sigma_{\text{exp}}
+ \Sigma_{\text{emul}}(\boldsymbol{\theta})
+ \Sigma_{\text{disc}}$. Here, $\Sigma_{\text{exp}}$ is the experimental covariance,  
$\Sigma_{\text{emul}}(\boldsymbol{\theta})$ is the GP emulator predictive covariance,  
and $\Sigma_{\text{disc}}$ is a (typically diagonal) model discrepancy term, and m represents the model's output dimensionality.

In practice, a realistic likelihood must also account for correlated sources of uncertainty. Experimental measurements are rarely statistically independent. Systematic uncertainties, in particular, often introduce correlations across centrality bins, momentum bins, and particle species. These correlations can significantly affect parameter inference; neglecting them typically leads to overly aggressive posteriors and artificially small credible intervals, etc. However, a persistent challenge in heavy-ion Bayesian calibration is that experimental collaborations do not make their full covariance matrices publicly available. Most published results provide point estimates with total uncertainties, occasionally decomposed into statistical and systematic components, but without the detailed bin-to-bin correlation structure required for a fully consistent probabilistic treatment. As a result, the experimental covariance $\Sigma_{\text{exp}}$ must be constructed or approximated. A common choice, which will be assumed here, is to treat the covariance matrix as uncorrelated (diagonal block). Ultimately, without full covariance matrices from the experiments, any Bayesian analysis must rely on informed but imperfect assumptions. Being explicit about these assumptions and transparent about their impact on parameter constraints is crucial for the credibility of the inference.  In future work, we will study the effects of correlated uncertainties on Bayesian inference. However, as described in the following sections, we note that they are not expected to affect the conclusions of this work.

Because evaluating \(\mathcal{M}(\boldsymbol{\theta})\) (the full simulator) is expensive, we replace it with a fast surrogate model (emulator) that provides a predictive mean \(\mu_\mathcal{M}(\boldsymbol{\theta})\) and predictive covariance \(\Sigma_{\text{emul}}(\boldsymbol{\theta})\).

Inspired by previous works, we employ the standard GP implementation from the Scikit-learn Python package. Another common practice is to perform a Principal Component Analysis (PCA) on the training data to reduce the number of required GPs. We included 6 principal components for all viscous correction models in Pb--Pb 2.76 TeV.

The emulator uncertainty must be included in the likelihood to avoid overconfident inference. Table~\ref{tb:prior_table} summarizes the model parameters and their ranges used in this work. We trained our emulators using 321 design points for the Grad model, 321 points for Chapman-Enskog, 319 points for PTM, and 318 points for PTB for Pb--Pb collisions at 2.76 TeV. The design points were all sampled using Latin Hypercube Sampling (LHS), as described in Ref.~\cite{JETSCAPE:2020mzn}. 
%
\begin{table*}
\footnotesize
\centering
\begin{tabular}{|p{0.24\linewidth}|p{0.12\linewidth}|p{0.12\linewidth}|p{0.24\linewidth}|p{0.12\linewidth}|p{0.10\linewidth}|}
\hline
\multicolumn{6}{|c|}{\textbf{Initial-state (5) and pre-equilibrium parameters (2)}} \\
\hline
Description & Parameter & Prior & Description & Parameter & Prior \\
\hline
Normalization & $N$[2.76 TeV] & {[}10, 20{]} &
Generalized mean & $p$ & {[}--0.7, 0.7{]} \\
Nucleon width & $w$ [fm] & {[}0.5, 1.5{]} &
Min. distance btw. nucleons & $d_{\text{min}}^3$ [fm$^3$] & {[}0, 1.7$^3${]} \\
Multiplicity fluctuation & $\sigma_k$ & {[}0.3, 2.0{]} &
Free-streaming time scale & $\tau_R$ [fm/$c$] & {[}0.3, 2.0{]} \\
Free-streaming energy dep. & $\alpha$ & {[}--0.3, 0.3{]} &
 &  &  \\
\hline
\multicolumn{6}{|c|}{\textbf{Hydrodynamics: QGP transport properties (9)}} \\
\hline
Description & Parameter & Prior & Description & Parameter & Prior \\
\hline
Temperature of $(\eta/s)$ kink & $T_{\eta}$ [GeV] & {[}0.13, 0.30{]} &
$(\eta/s)$ at kink & $(\eta/s)_{\rm kink}$ & {[}0.01, 0.20{]} \\
Low-$T$ slope of $(\eta/s)$ & $a_{\text{low}}$ [GeV$^{-1}$] & {[}--2, 1{]} &
High-$T$ slope of $(\eta/s)$ & $a_{\text{high}}$ [GeV$^{-1}$] & {[}--1, 2{]} \\
Shear relaxation factor & $b_{\pi}$ & {[}2, 8{]} &
Max. of $(\zeta/s)$ & $(\zeta/s)_{\text{max}}$ & {[}0.01, 0.25{]} \\
Temperature of $(\zeta/s)$ peak & $T_{\zeta}$ [GeV] & {[}0.12, 0.30{]} &
Width of $(\zeta/s)$ peak & $w_{\zeta}$ [GeV] & {[}0.025, 0.15{]} \\
Asymmetry of $(\zeta/s)$ peak & $\lambda_{\zeta}$ & {[}--0.8, 0.8{]} &
 &  &  \\
\hline
\multicolumn{6}{|c|}{\textbf{Particlization (1)}} \\
\hline
Description & Parameter & Prior & \multicolumn{3}{c|}{} \\
\hline
Switching temperature & $\Tsw$ [GeV] & {[}0.135, 0.165{]} &
\multicolumn{3}{c|}{} \\
\hline
\end{tabular}
\caption{17 model parameters and prior ranges from Ref.~\cite{JETSCAPE:2020mzn}. Parameters are grouped according to the stage of the collision where they predominantly influence the dynamics. The prior is uniform within the quoted ranges.}
\label{tb:prior_table}
\end{table*}
%
Each likelihood evaluation may call the emulator (fast surrogate model), but it could still be costly if \(\Sigma_{\text{tot}}(\boldsymbol{\theta})\) is large and inversion is needed; naive MCMC may require \(10^4\!-\!10^6\) evaluations. Parallel tempering ensemble sampling (ptemcee) helps explore multimodal posteriors and improves mixing. It is used to draw samples \(\{\boldsymbol{\theta}^{(s)}\}_{s=1}^S\) from \(p(\boldsymbol{\theta}\mid D)\). In this work, we sample the posterior using the affine‐invariant MCMC ensemble sampler \texttt{emcee}~\cite{Foreman-Mackey:2012any} with 200 walkers and \(5\times10^4\) steps, discarding the first \(10^4\) steps as burn‐in.
\subsection{Global Sensitivity Analysis}
\label{subsec:GSA}
\begin{figure}
    \centering
    \includegraphics[width=\linewidth]{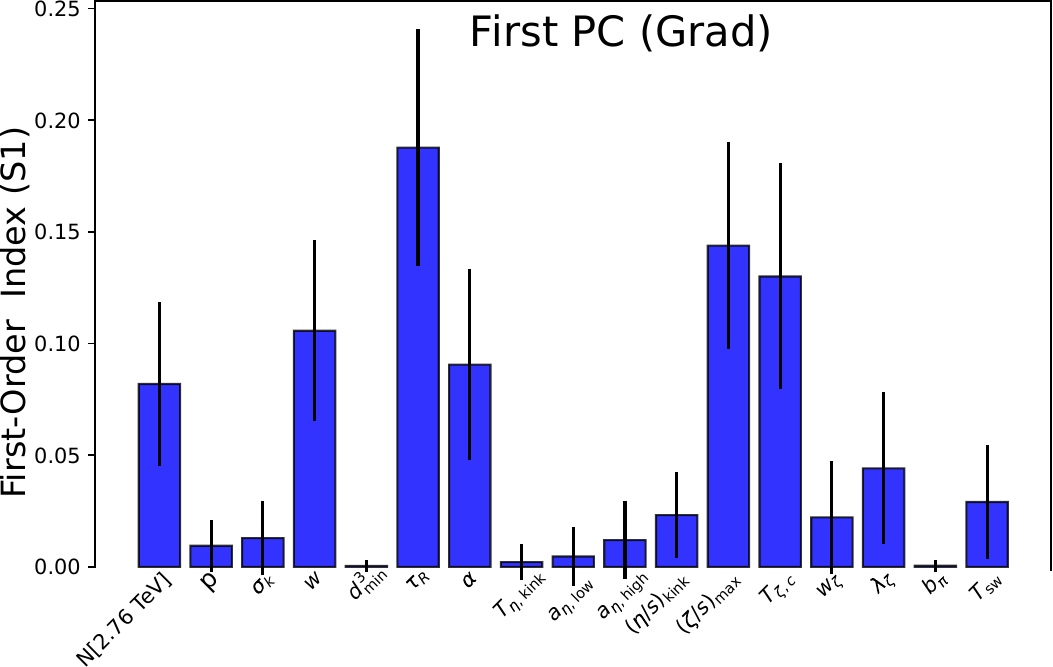}
    \caption{First-order Sobol sensitivity indices for the scaled spectra $U(x_T)$, computed using Gaussian process emulators for the Grad viscous correction. The analysis identifies the free-streaming time scale $\tau_R$, the maximum bulk viscosity $(\zeta/s)_{\rm max}$, the position of the peak of the bulk viscosity $T_{\zeta}$, and the Gaussian nucleon width $w$ as the dominant parameters controlling the overall shape and scaling behavior of the spectra. Their combined influence underscores the role of pre-equilibrium dynamics and initial-state granularity in establishing the observed universality.}
    \label{fig:Grad_sobol_index}
\end{figure}

A critical step in our Bayesian workflow is to identify which model parameters most strongly influence the shape of the scaled spectrum \(U(x_T)\) and \emph{the constraining power} of this proposed observable. Through a global sensitivity analysis (GSA), we can quantify how strongly our model predictions depend on specific model parameters, disentangle correlations, and quantify degeneracies over the parameter space~\cite{Paquet:2023rfd}.
To this purpose, we explore a global measure to quantify this information; we use Sobol's indices to quantify the global variance of model observables (model output) corresponding to the variances in the model parameters (model inputs). 
We can define a first-order Sobol index for a particular observable $y$ and a specific model parameter \(\boldsymbol{\theta}_i\) according to 
\begin{equation}
S_i \;=\; \frac{\mathrm{Var}\bigl[\mathbb{E}(y\mid\boldsymbol{\theta}_i)\bigr]}{\mathrm{Var}(y)},    
\label{eq:sobol_index}
\end{equation}
this quantity represents the variance in the observable \(y\) corresponding to the variance of the parameter \(\boldsymbol{\theta}_i\).
Using Saltelli’s sampling scheme~\cite{Saltelli2002}, we compute the first‐order Sobol index for each input parameter \(\boldsymbol{\theta}_i\), where \(y\) denotes a principal‐component score of \(U(x_T)\).  We perform this analysis for the first leading principal component (PC1), which captures over 80\% of the variance in the scaled spectrum (see Appendix Fig.~\ref{fig:explained_variance_Grad}). In our analysis, the Sobol indices among principal component scores \(z_\alpha\) and model parameters \(\boldsymbol{\theta}_i\) were estimated using the Python library SALib~\cite{Iwanaga2022, Herman2017} and are shown in Fig.~\ref{fig:Grad_sobol_index} for the first principal component of the Grad model.  
\section{Results}
\label{sec:results}
In this section, we present the main outcomes of our Bayesian analysis, organized into four parts. First, we compare the prior and posterior ensembles of model predictions to experimental data. We then discuss the global sensitivity analysis that identifies the most influential parameters, ranking them using Sobol's index. This analysis provides interpretative power to the results of the posterior distribution. Finally, we compare the four different viscous correction models, Sec.~\ref{subsec:particlization}, discussing their effects on the shape of the spectra. 
\subsection{MODEL PREDICTIONS COMPARED TO EXPERIMENTAL DATA}
\label{subsec:model_vs_data}
\begin{figure}[ht]
  \centering
  \includegraphics[width=\linewidth]{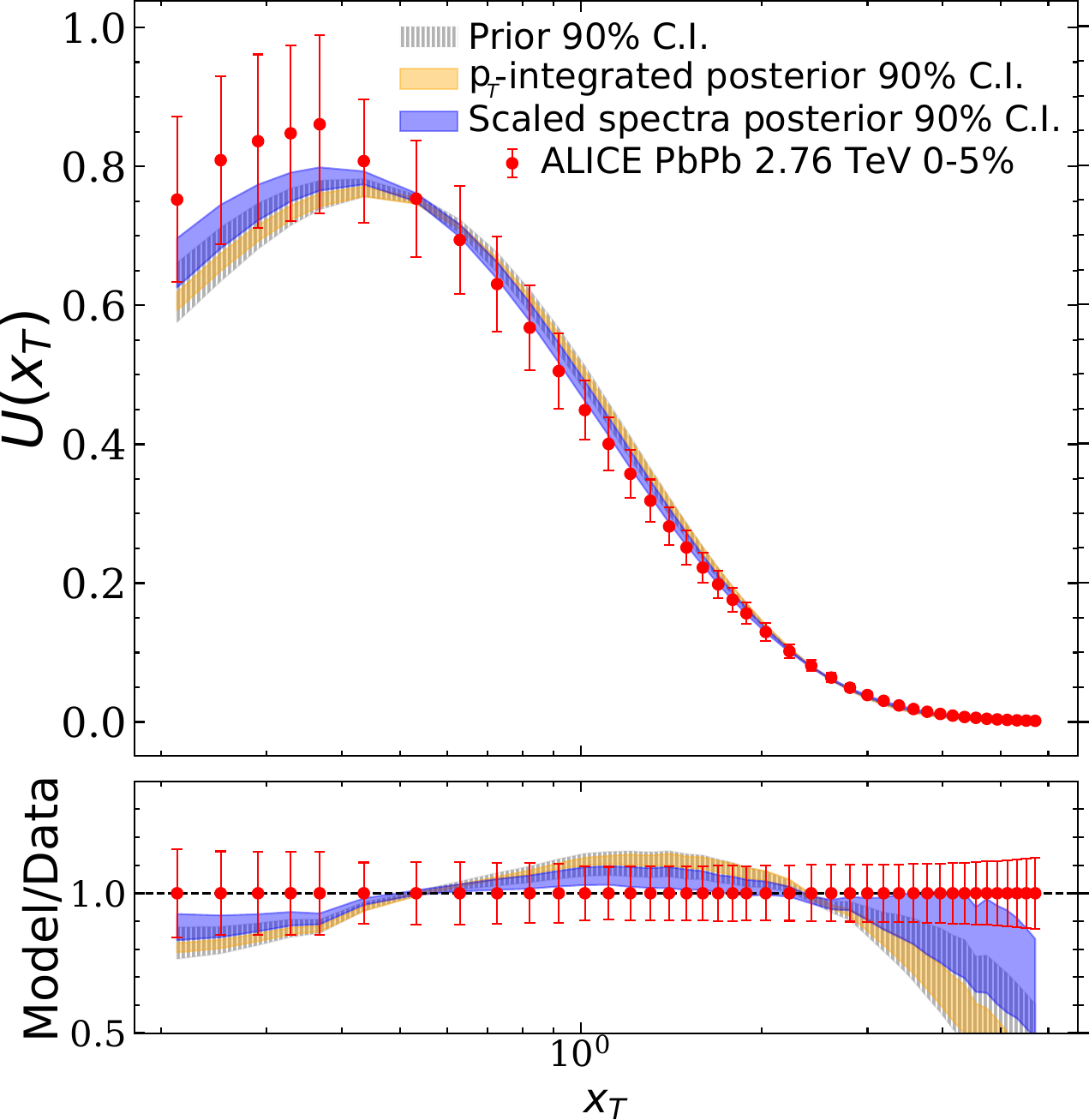}
  \caption{Prior (gray hatched band) and posterior (orange and blue bands) model predictions for the scaled spectra $U(x_T)$ obtained with the Grad viscous correction, compared with ALICE Pb--Pb data at $\sqrt{s_{NN}} = 2.76~\mathrm{TeV}$ for the most central events. The narrow prior bands reflect the intrinsic universality of $U(x_T)$, while the posterior 
  calibration sharpens these predictions, capturing the overall shape but revealing systematic deviations at low and high $x_T$. Experimental data points include combined uncorrelated statistical and systematic uncertainties. The original JETSCAPE posterior~\cite{JETSCAPE:2020mzn} calibrated to $p_T$-integrated observables are displayed by the orange bands. The scaled-spectra posterior, represented by the blue band, is consistent with data but differs systematically from the integrated-observable posterior, indicating a tension between the two calibration strategies and highlighting the sensitivity of $U(x_T)$ to distinct regions of parameter space.}
  \label{fig:model_vs_data}
\end{figure}
The JETSCAPE framework provides a flexible, multistage description of relativistic heavy-ion collisions~\cite{Putschke:2019yrg}, with model parameters controlling various aspects of the initial conditions, pre-equilibrium evolution, transport coefficients, and particlization. In this work, we employ Bayesian parameter estimation (BPE) to investigate how well this state-of-the-art model reproduces the experimentally observed universality of the scaled particle spectra for pions, $U(x_T)$, and to assess the constraining power of this observable relative to traditional $p_T$-integrated quantities.

The prior distribution defines the initial range of physically plausible model parameters. In~\cite{JETSCAPE:2020mzn}, they used 17 parameters for Pb--Pb 2.76 TeV, spanning a broad space of initial conditions and transport model features to ensure full coverage of the experimentally accessible observables. The priors were originally designed to reproduce the soft sector and $p_T$-integrated observables, such as charged-particle yields, identified mean transverse momenta, fluctuations, and anisotropic flow coefficients.

When applied to the scaled spectra $U(x_T)$, depicted in Fig.~\ref{fig:model_vs_data}, the prior predictions show relatively narrow 90\% credible intervals, indicating that the model, even before calibration, exhibits limited flexibility in reproducing this observable. The narrow prior band reflects the intrinsic universality of $U(x_T)$ observed in both experimental data and hydrodynamic model predictions. In other words, the scaled spectra appear to be a robust emergent property of the hydrodynamic evolution, insensitive to large variations in individual model parameters within the prior range.

A calibration of only the scaled spectra produces a posterior distribution (represented by the blue credible band in Fig.~\ref{fig:model_vs_data}) that significantly sharpens relative to the prior, revealing the constraining power of $U(x_T)$. Overall, the posterior predictions appear consistent with experimental data within uncertainties, although we caution that the exact quality of the fit depends on whether and how the experimental errors are correlated. Nevertheless, the general tendency is to underestimate the spectra at small $x_T$, overshoot the data at intermediate $x_T$, and again underestimate at large $x_T$, as depicted by the panel below representing the model to data ratio. 

A centrality dependence is also observed when we compare Fig.~\ref{fig:model_vs_data} with Fig.~\ref{fig:Universal_viscous_30-40}(a): the posterior bands are noticeably broader for more peripheral collisions, likely due to reduced event statistics and weaker collective behavior in these systems. In contrast, the predictions for central events are more tightly constrained. 

We next compare the scaled spectra posterior with the original JETSCAPE result (orange band)~\cite{JETSCAPE:2020mzn}, which was obtained from a calibration using only $p_T$-integrated observables. These include the charged and identified particle yields $dN/dy$, $\langle  p_T \rangle$, $p_T$ fluctuations $\delta p_T / \langle p_T \rangle$, and the integrated anisotropic flow coefficients $v_n\{2\}$ for $n=2,3, 4$. This comparison serves as a validation test: a consistent Bayesian model should describe observables not used in the calibration. However, we find that the two posteriors are inconsistent. The scaled-spectra posterior and the original posterior prefer distinct regions of the parameter space, highlighting a tension between the model’s ability to simultaneously describe integrated and differential observables. A natural question is \textit{where this tension comes from}. 
\begin{figure}
    \centering
    \includegraphics[width=\linewidth]{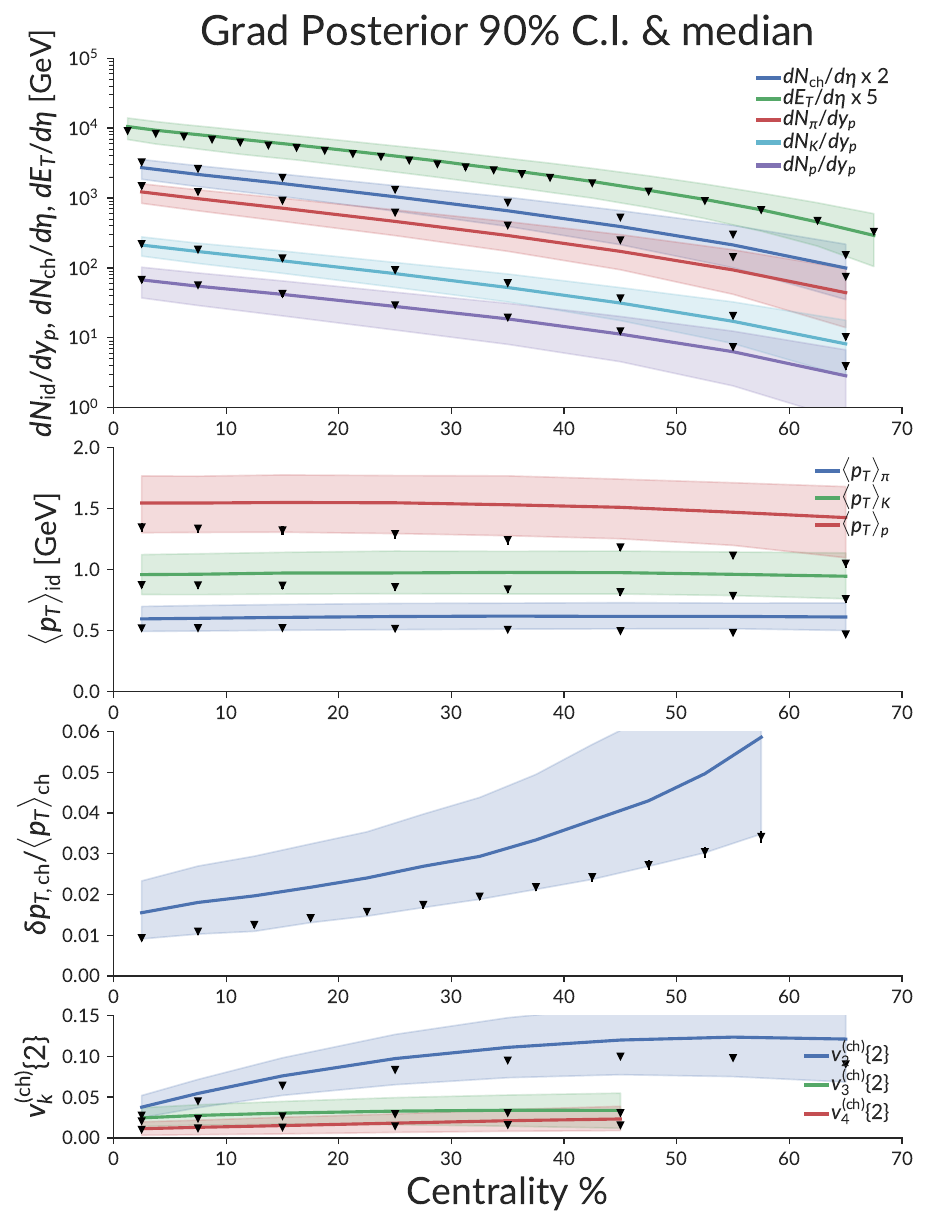}
    \caption{Posterior predictions (median and 90\% credible intervals) from the {\it scaled-spectra} calibration using the Grad viscous correction, shown for \emph{\(p_T\)-integrated} observables in Pb--Pb collisions at \(\sqrt{s_{NN}}=2.76\)~TeV.  The posterior reproduces the general centrality trends of particle yields, mean transverse momenta, \(p_T\) fluctuations, and anisotropic flow coefficients \(v_n\{2\}\), but systematic deviations appear in the mean \(\langle p_T\rangle\) and its fluctuations. These discrepancies indicate the tension between the calibration to the scaled spectra \(U(x_T)\) and the traditional \(p_T\)-integrated observables. Black triangles represent ALICE experimental data~\cite{ALICE:2010mlf, ALICE:2016igk, ALICE:2013mez, ALICE:2011ab, ALICE:2014gvd}.}
    \label{fig:integrated_observables_spectra_posterior}
\end{figure}
A closer examination reveals that the tension arises primarily from the mean transverse momentum of identified particles $\langle p_T \rangle$ and its variance; see Fig.~\ref{fig:integrated_observables_spectra_posterior}. The predictions for $p_T$-integrated observables using the scaled spectra calibration tend to overestimate the mean $p_T$ relative to ALICE experimental data, leading to systematic deviations when extrapolated to $U(x_T)$.

%
\subsection{GLOBAL SENSITIVITY ANALYSIS AND PARAMETER POSTERIORS}
\label{subsec:sensitivity_analysis_params}
To identify which physical parameters most strongly influence the shape of the scaled spectra $U(x_T)$, we performed a Global Sensitivity Analysis (GSA) using the trained Gaussian process emulators within the JETSCAPE Bayesian framework. The Sobol index, described by Eq.\eqref{eq:sobol_index}, quantifies the relative importance of each model parameter in controlling the model output, allowing us to rank the dominant contributors to the universal shape of $U(x_T)$.
The GSA reveals that four parameters dominate the response of $U(x_T)$:
\begin{itemize}
    \item The free-streaming time scale $\tau_R$, which controls the pre-equilibrium dynamics and sets the initial conditions for hydrodynamic evolution;
    \item The maximum bulk viscosity $(\zeta/s)_{\rm max}$, which regulates entropy production and collective expansion near the QCD crossover region;
    \item The temperature of the bulk viscosity  peak $T_{\zeta}$, i.e., the temperature at which we assume that the bulk viscosity has a single peak, is near the deconfinement temperature;
    \item The Gaussian nucleon width $w$, representing the effective transverse size of the nucleons and determining the granularity of the initial density profile.
\end{itemize}
These parameters collectively govern the overall shape and scaling behavior of $U(x_T)$. Variations in $\tau_R$, $(\zeta/s)_{\rm max}$, and $T_{\zeta}$ modify the global curvature of the spectra, while the nucleon width 
$w$ directly affects its steepness and degree of universality across different collision systems and centralities. Their prominent sensitivity underscores that the universal behavior of the scaled spectra is not an accidental feature, but rather a controlled outcome of specific aspects of the pre-equilibrium and initial-state physics.

\begin{figure}
    \centering
    \includegraphics[width=\linewidth]{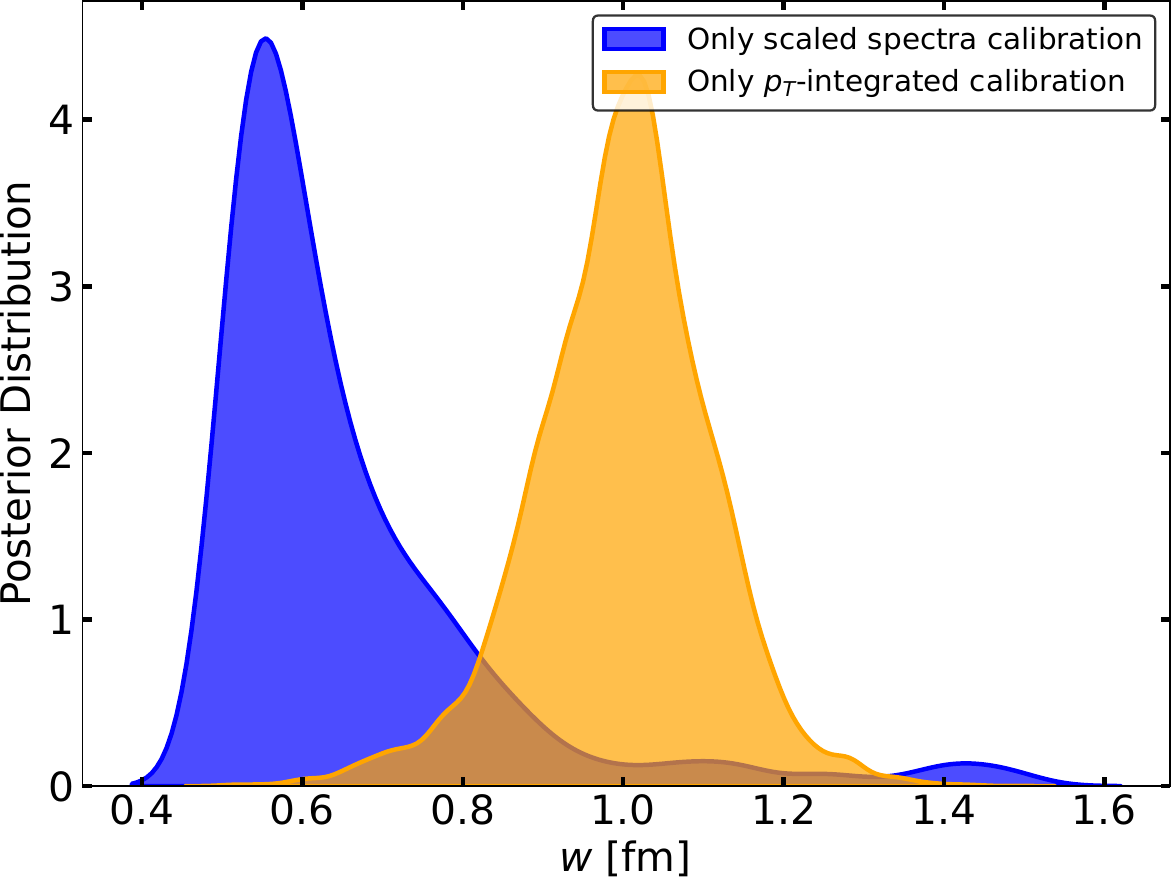}
    \caption{Posterior distributions of the Gaussian nucleon width $w$ from Bayesian calibrations using scaled-spectra data (blue) and $p_T$-integrated observables (orange). The scaled-spectra posterior favors smaller values of $w \sim 0.6~\mathrm{fm}$, corresponding to more granular initial conditions, whereas the $p_T$-integrated calibration prefers smoother initial profiles with $w \sim 1.0~\mathrm{fm}$. This opposing behavior highlights the origin of the tension between the two calibration approaches.}
    \label{fig:w_posterior}
\end{figure}

A direct comparison between the posterior distributions of the Gaussian nucleon width $w$ from different Bayesian calibrations, shown in Fig.~\ref{fig:w_posterior}, further illustrates the underlying tension among observables.
The posterior obtained from the scaled-spectra calibration (blue distribution) strongly favors smaller values of $w$, around $0.6$ fm, indicating a preference for more granular initial conditions, where nucleons are effectively compact and localized. In contrast, the posterior derived from the $p_T$-integrated calibration—which includes yields, $\langle p_T \rangle$, fluctuations, and integrated flow coefficients—prefers larger widths, approximately $1.0$ fm, corresponding to smoother initial density profiles typically favored in previous heavy-ion Bayesian analyses~\cite{JETSCAPE:2020shq, Bernhard:2019bmu}.

This divergence reveals a clear and physically interpretable tension:
the scaled-spectrum observable $U(x_T)$ favors smaller nucleon widths and thus more granular structures in the initial conditions, whereas the $p_T$-integrated observables, especially the mean transverse momentum, are better reproduced with smoother initial profiles. Consequently, $U(x_T)$ and $\langle p_T \rangle$ probe distinct aspects of the initial-state geometry, emphasizing the need for a joint calibration strategy capable of reconciling these competing constraints.

Thus, the scaled spectra $U(x_T)$ join other observables such as the total nucleon--nucleon cross section \cite{Nijs:2022rme} and $\rho_2$ (the correlation between $\langle p_T\rangle$ and $v_2$) \cite{Giacalone:2021clp} as an observable that requires small values for $w$.

This inability to simultaneously describe multiple observables (for any combination of model parameters) indicates that something is likely missing in the model and motivates further research to identify any missing physics that would resolve this tension.  For example, it has been suggested that non-equilibrium Goldstone mode dynamics should enhance the yield of pions at low transverse momentum compared to calculations that neglect this physics \cite{Florio:2025zqv, Florio:2025lvu}.   This enhancement would improve the fit to the scaled spectra and could potentially resolve the model tension.
\subsection{COMPARING DIFFERENT VISCOUS CORRECTIONS}
\label{subsec:viscous_corrections}
\begin{figure*}[ht!]
    \centering
    
    \begin{subfigure}[b]{0.48\textwidth}
        \centering
        \includegraphics[width=\linewidth]{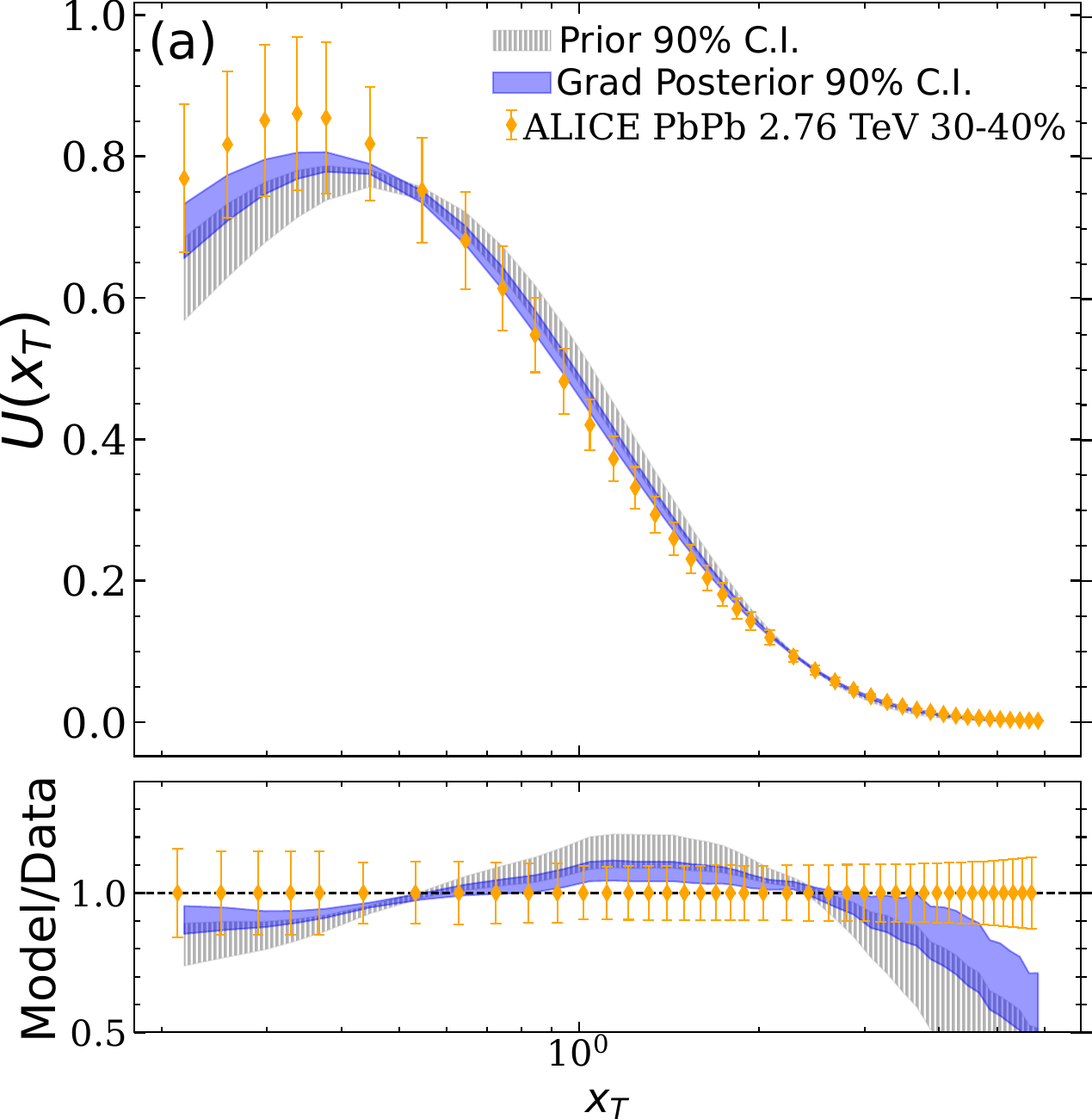}
        \label{fig:Universal_Grad_30-40}
    \end{subfigure}
    \begin{subfigure}[b]{0.48\textwidth}
        \centering
        \includegraphics[width=\linewidth]{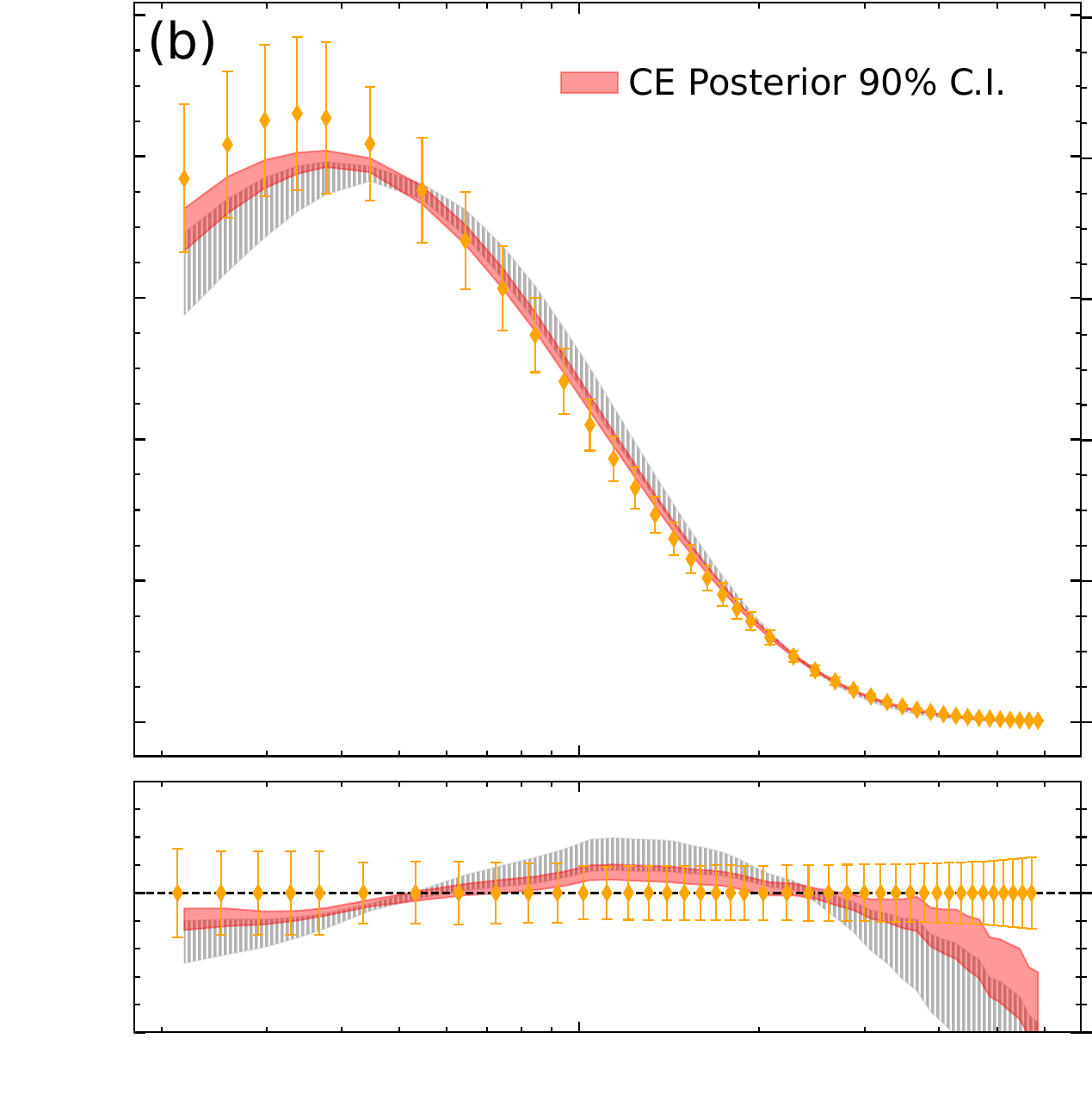}
        \label{fig:Universal_CE_30-40}
    \end{subfigure}

    \begin{subfigure}[b]{0.48\textwidth}
        \centering
        \includegraphics[width=\linewidth]{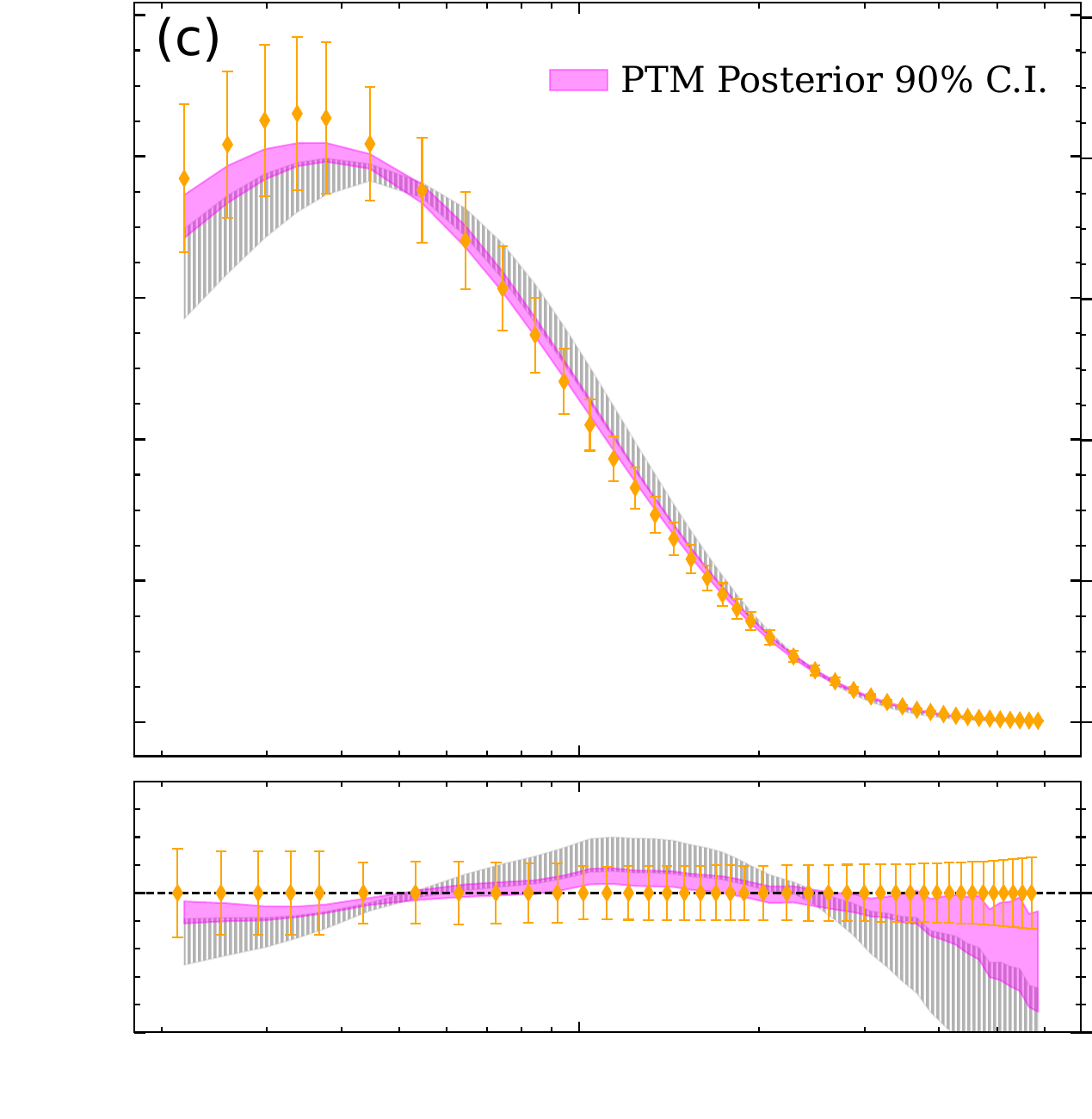}
        \label{fig:Universal_PTM_30-40}
    \end{subfigure}
    \begin{subfigure}[b]{0.48\textwidth}
        \centering
        \includegraphics[width=\linewidth]{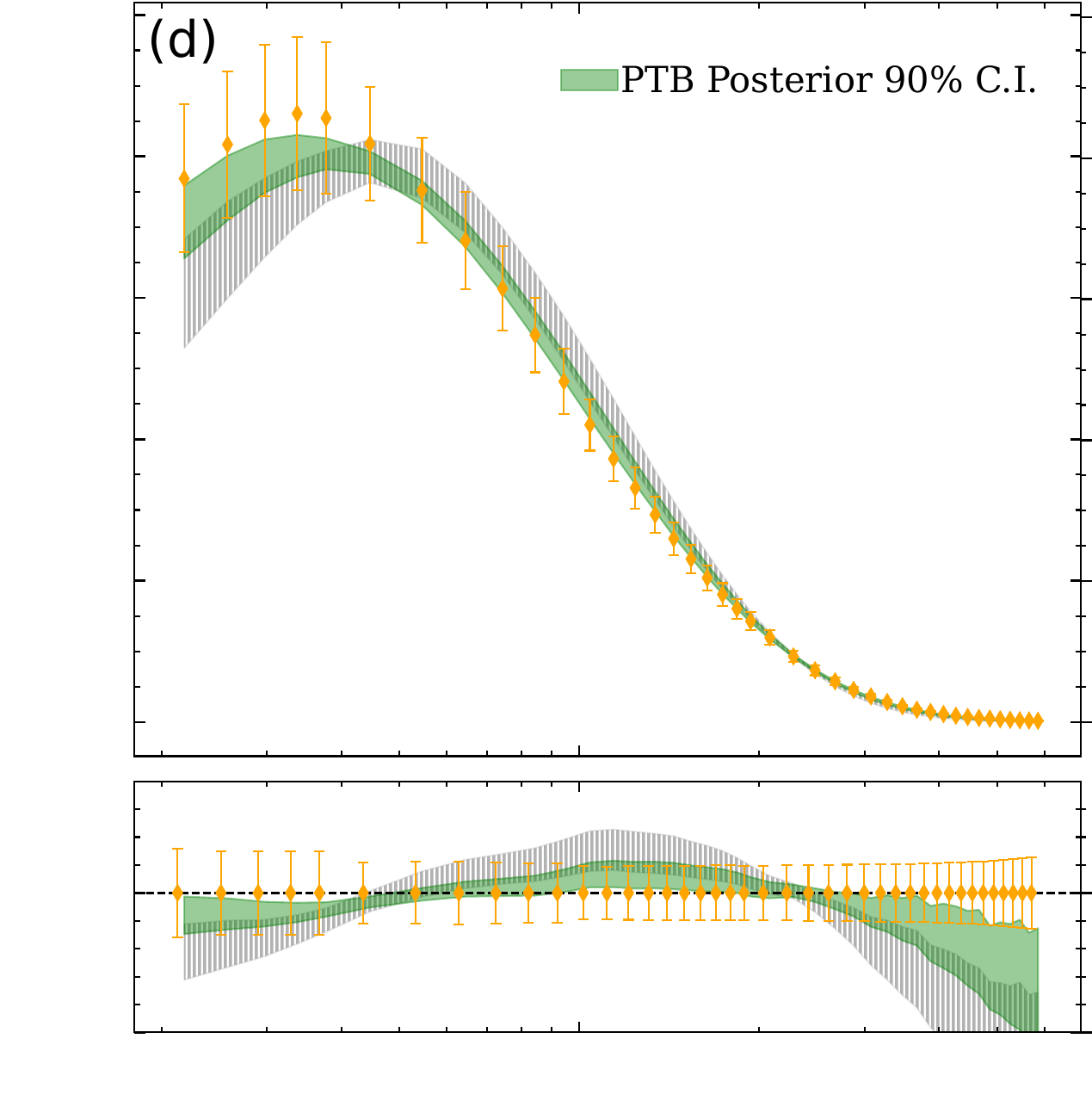}
        \label{fig:Universal_PTB_30-40}
    \end{subfigure}
    \caption{
    Model-to-data comparison of the scaled spectra $U(x_T)$ for 30-40\% centrality using four different viscous correction prescriptions at particlization: Grad (a), Chapman--Enskog (CE) (b), Pratt--Torrieri--McNelis (PTM) (c), and Pratt--Torrieri--Bernhard (PTB) (d).
    While all models preserve the overall universality of $U(x_T)$, the PTB implementation exhibits noticeably larger posterior uncertainty and stronger sensitivity to bulk viscous parameters~(see Fig.~\ref{fig:GSA_PTB}), reflecting the nonlinear structure of its exponentiated viscous correction.
    In contrast, the linear Grad, CE, and PTM corrections display comparable behavior and a shared sensitivity to both the bulk viscosity peak $(\zeta/s)_{\rm max}$ and the free-streaming relaxation time $\tau_R$ (see Appendix section~\ref{appendix:GSA_viscous_corrections}).
    These results demonstrate that the universal scaling of $p_T$ spectra is primarily governed by the collective expansion dynamics and remains robust under variations of the particlization scheme.
    }
    \label{fig:Universal_viscous_30-40}
\end{figure*}
An additional source of theoretical uncertainty in hydrodynamic modeling arises from the choice of viscous correction, and an interesting question could be how the different models affect the universal spectra shape. As mentioned in Sec.~\ref{subsec:particlization}, 
each of the models embodies distinct assumptions about the nonequilibrium distribution of hadrons and the mapping between fluid-dynamical and kinetic degrees of freedom at freeze-out. In Fig.~\ref{fig:Universal_viscous_30-40}, we show a comparison between the 4 viscous corrections for a specific centrality bin. Among the four implementations, the PTB correction, shown in panel d, stands out as the most distinct. The PTB formulation is based on an exponentiated ansatz for the viscous correction, as opposed to the linearized Grad and CE forms.
While this functional structure ensures the physical positivity of the phase-space distribution, it also introduces nonlinear dependencies on the viscous fields that complicate emulator training and increase predictive uncertainty. Indeed, the PTB model exhibits noticeably larger posterior uncertainties and lower emulator accuracy, as discussed in detail in the emulator validation section of the Appendix~(\ref{appendix:emulator_PCA}).

Despite these differences, the scaled particle spectra $U(x_T)$ show remarkable robustness with respect to the choice of viscous correction. The overall shape and degree of universality are only weakly affected by the specific $\delta f$ implementation. This indicates that the scaled spectra are predominantly governed by the initial conditions and bulk dynamical evolution of the system, rather than by the details of the particlization procedure. Nevertheless, the PTB correction displays stronger sensitivity to the bulk viscosity parameters, particularly to the maximum value 
$(\zeta/s)_{\rm max}$, compared with the other formulations. In contrast, the Grad, CE, and PTM corrections exhibit a combined sensitivity to both the free-streaming relaxation time $\tau_R$ and $(\zeta/s)_{\rm max}$. This difference suggests that the exponentiated form of PTB tends to amplify the role of bulk viscous effects while reducing the influence of early-time pre-equilibrium dynamics.

While in principle, one could obtain almost any $p_T$-dependence by sufficiently tuning the functional form of $\delta f$, these comparisons  reinforce the conclusion that for reasonable choices of $\delta f$, the universal behavior of $U(x_T)$ appears to be an emergent property of the collective expansion dynamics and is largely insensitive to moderate variations in the particlization scheme.
\section{CONCLUSIONS AND DISCUSSION}
\label{sec:conclusions}
In this work, we explored the scaled pion $p_T$ spectra,  $U(x_T)$, within the framework of Bayesian parameter estimation applied to a state-of-the-art \trento{}+hydro+afterburner model for relativistic heavy-ion collisions.
By combining emulation techniques, global sensitivity analysis, and Bayesian calibration, we systematically investigated the constraining power of the scaled spectra, their sensitivity to model parameters, and the physical insights that can be extracted from their near-universal behavior across systems and centralities.

Our results demonstrate that, despite the success of the hybrid model framework in describing a wide variety of soft observables, current state-of-the-art models struggle to reproduce the precise universal shape of $U(x_T)$ observed experimentally.
Even after calibration, the model barely agrees within a 1-sigma interval (and may, in fact, disagree depending on how the experimental errors are correlated).  Generally, the results tend to underestimate the data at low $x_T$, overestimate around intermediate $x_T$, and again underestimate at higher $x_T$.
A sensitivity analysis reveals that the observable is predominantly influenced by four key parameters: the free-streaming time scale $\tau_R$, the maximum bulk viscosity 
$(\zeta/s)_{\rm max}$, and the Gaussian nucleon width $w$. These parameters jointly determine the shape and degree of universality of the spectra, controlling both the early-time smoothing of the initial energy density and the subsequent collective expansion of the medium. 

However, a significant tension arises between the $U(x_T)$-based calibration and the posterior derived from $p_T$-integrated observables, particularly the mean transverse momentum and its fluctuations. While the scaled spectra favor smaller nucleon widths ($w\sim 0.6$ fm) — corresponding to more granular initial conditions — the mean transverse momentum is better described by smoother profiles ($w \sim 1.0$ fm). This contrast highlights that $U(x_T)$ and $\langle p_T \rangle$ may probe complementary aspects of the system: the former constrains the spectral shape and collective response to initial granularity, while the latter reflects the average collective flow and temperature gradients.

These findings suggest that the apparent universality of $U(x_T)$ is not accidental but emerges from the combined effects of initial-state smoothing and collective hydrodynamic evolution. Finally, our results indicate that the universal shape of $U(x_T)$ is relatively insensitive to the specific viscous correction model used in the particlization stage.
Among the models studied, the Pratt–Torrieri–Bernhard (PTB) correction — based on an exponentiated viscous ansatz — shows the largest deviation from the Grad and Chapman–Enskog (CE) linearized corrections, though the overall impact on the scaled spectra remains modest. 

In summary, this work establishes the scaled spectra $U(x_T)$ as a robust and complementary observable for Bayesian analyses of heavy-ion collisions. It provides unique sensitivity to parameters governing the early-time and initial-state dynamics, offering a novel perspective on the emergence and limits of hydrodynamic behavior in strongly interacting matter. 

In particular, the inability to simultaneously describe the scaled spectra along with $p_T$-integrated observables (for any combination of model parameters) likely indicates missing physics in the model and motivates further study.   For example, one could speculate that missing low-$p_T$ pions from non-equilibrium Goldstone mode dynamics \cite{Florio:2025zqv, Florio:2025lvu} may potentially explain the discrepancy.
\section*{Acknowledgments}
We thank the JETSCAPE Collaboration for access to their hybrid simulation framework, Bayesian analysis tools, and for sharing simulation data; we also thank the ALICE Collaboration for making their data publicly available.
This work was supported by FAPESP project 2023/13749-1, CNPq project 405458/2025-8, and INCT-FNA grant under process No. 408419/2024-5.   
The author T.S.D. acknowledges financial support from the Brazilian National Council for Scientific and Technological Development (CNPq) through a PhD fellowship under process No.\ 141432/2025-0, INCT-FNA grant number 464898/2014-5, and from the Brazilian Federal Agency for Support and Evaluation of Graduate Education CAPES Foundation scholarship under process No. 88881.220538/2025-01. M.L. was supported by FAPESP projects 2017/05685-2 and 2018/24720-6, and by project INCT-FNA Proc. No. 464898/2014-5. T. N.dS. is supported by CNPq through the INCT-FNA grant 312932/2018-9 and the Universal Grant 409029/2021-1. C.D.M.~was supported by FAPESP projects 2023/13749-1 and 2025/01122-0. J.T.~was supported by FAPESP project 2023/13749-1 and by CNPq through 303650/2025-7. G. S. D. acknowledges support from CNPq. F.G.G. was supported by CNPq through 307806/2025-1 and 405458/2025-8.
M.H. was supported by CNPq under process No. 313638/2025-0. 
\section*{Data Availability}

This study makes use of both theoretical predictions generated by the authors and publicly available experimental data. 
The simulated datasets, together with the trained emulators, design points, parameter ranges, experimental data, posterior distributions, and analysis scripts used in this work, are openly available in the GitHub repository associated with this manuscript~\cite{DominguesUniversalityRepo2025}.
The repository also provides a detailed description of the Bayesian analysis workflow, global sensitivity analysis, further investigations, and an interactive \texttt{streamlit} application for exploring model predictions.

All experimental data used in this analysis are stored in the same GitHub repository and are also available from the corresponding published sources and their associated repositories (e.g., HEPData), as cited throughout the manuscript.
\appendix
\label{sec:appendix}
\section{Principal Component Analysis of Scaled Spectra}
\label{appendix:appendix_pca}
Given the inherent high dimensionality of the scaled spectrum observable $U(x_T)$, which is characterized by 41 discrete $x_T$ bins across 7 distinct centrality bins, the application of Principal Component Analysis (PCA) becomes a crucial step. By transforming the data into a new set of orthogonal variables, called principal components, PCA allows us to capture the most significant variations in the data with a much smaller number of dimensions, thus simplifying the subsequent analysis and model training.

We organized the model output training data into a matrix $\mathbf{X}\in\mathbb{R}^{N_{\rm dp}\times m}$. In this matrix, each row corresponds to a unique design point, representing a specific configuration within our parameter space, and each column corresponds to the value of $U(x_T)$ at a particular bin, effectively capturing the spectrum's behavior across its various bins in $x_T$ and centrality for that design point. To ensure that our analysis focuses purely on variability and not on the absolute magnitudes, each column is centered by subtracting its mean value and dividing by the covariance over the entire design space, resulting in zero mean and unit covariance data. This standardization step prevents features with larger scales from disproportionately influencing the principal components.

Following the standardization of the data, we computed the empirical covariance matrix $\mathbf{C}$. This matrix is defined as
\begin{equation}
\mathbf{C} = \frac{1}{N_{\rm dp}-1}\,\mathbf{X}^\mathsf{T}\mathbf{X}\,,    
\label{eq:covariance_matrix_appendix}
\end{equation}
which quantifies the relationships and variances between all pairs of the $m$ model output bins. Dividing by $N_{\rm dp}-1$ yields an unbiased estimate of the population covariance. The eigenvectors of this covariance matrix represent the principal components (PCs), which are the directions of maximum variance in the data. The eigenvalues associated with these eigenvectors indicate the magnitude of variance along each respective principal component. Specifically, we solve the eigenvalue problem $\mathbf{C}\,\mathbf{e}_\alpha = \lambda_\alpha\,\mathbf{e}_\alpha$ to obtain the ordered eigenvalues $\lambda_1\ge\lambda_2\ge\cdots\ge0$ and their corresponding orthonormal eigenvectors $\{\mathbf{e}_\alpha\}$. The principal components are ordered by their eigenvalues, with $\lambda_1$ representing the largest variance and so forth.

\begin{figure}[ht]
  \centering
  \includegraphics[width=\linewidth]{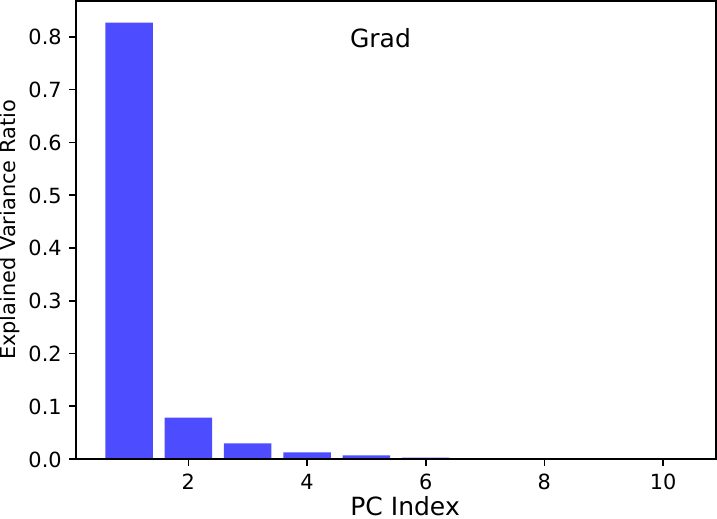}
  \caption{Principal Component Analysis (PCA) of Scaled Spectra. This figure illustrates the proportion of total variance explained by the first 10 principal components for the Grad viscous correction. The large amount of explained variance in the first principal component demonstrates a high correlation between all the $U(x_T)$ bins. This behavior is similar for the other off-equilibrium models (see associated figures stored in~\cite{DominguesUniversalityRepo2025}).}
  \label{fig:explained_variance_Grad}
\end{figure}

A key metric in PCA is the fraction of total variance explained by the first $K$ components, calculated as
\begin{equation}
\mathrm{Var}(K) = \frac{\sum_{\alpha=1}^K \lambda_\alpha}{\sum_{\alpha=1}^m \lambda_\alpha}\,.
\label{eq:frac_of_total_variance}   
\end{equation}
This metric provides a clear indication of how much of the original data's variability is retained when we project it onto a reduced subspace spanned by the first $K$ principal components. A higher $\mathrm{Var}(K)$ implies that a significant amount of information is preserved with fewer dimensions. We find that the first \(K=6\) principal components already capture \(>98\%\) of the variance for all viscous corrections, as shown in Fig.~\ref{fig:explained_variance_Grad}, which depicts the explained variance of each principal component for the Grad model. 
\section{Emulator construction in PCA Space and validation}
\label{appendix:emulator_PCA}
Instead of emulating each \(U(x_T)\) bin separately, we project the training data onto the first \(K\) principal components:
\begin{equation}
\mathbf{z}_k = \mathbf{X}_k \,\mathbf{E}_K,    
\label{eq:projected_training_data}
\end{equation}
where \(\mathbf{E}_K = [\mathbf{e}_1,\dots,\mathbf{e}_K]\).  We then build independent GP emulators for each principal‐component coefficient \(z_\alpha\). At prediction time, the emulator provides \(\hat z_\alpha(\boldsymbol{\theta})\), and the full spectrum is reconstructed as
\begin{equation}
\widehat{U}(x_T) \;=\; \bar U(x_T) \;+\; \sum_{\alpha=1}^K \hat z_\alpha(\boldsymbol{\theta})\,e_\alpha(x_T)\,,    
\label{eq:predictions_reconstruction}
\end{equation}
where \(\bar U(x_T)\) is the mean spectrum over the design. This PCA‐based strategy reduces the emulator dimensionality from \(m=287\) to \(K=6\), significantly improving training efficiency without compromising predictive accuracy.

To verify that our emulators faithfully reproduce the full hybrid model outputs, we performed a leave–one–out cross-validation (LOOCV)~\cite{loocv_scikitlearn} on each of our four viscous correction prescriptions (Grad, CE, PTM, PTB). The steps for each emulator's validation are:
\begin{enumerate}
  \item For $l=1,\dots,N_{\rm dp}$, remove the $l$‑th design point $\boldsymbol\theta_l$ from the training set of size $N_{\rm dp}-1$.
  \item Train the GP emulator on the remaining $N_{\rm dp}-1$ runs, using a composite kernel $C\cdot\mathrm{RBF}+ \mathrm{White}$ with optimized hyperparameters.
  \item Predict the full scaled spectrum $\hat U^{\rm GP}(x_T\,|\,\boldsymbol\theta_l)$ at the held‑out point and compare it to the direct simulation $U(x_T\,|\,\boldsymbol\theta_l)$.
  \item Record the Pearson correlation coefficient $r_l$ between true and predicted values over all $x_T$ bins, and accumulate the scatter.
\end{enumerate}
We assemble the $\{\,\hat U,\;U\}$ scatter for each held‑out point into a single plot per prescription.  All four emulators achieve $r>0.99$ across all centralities and $x_T$ bins, and the mean squared error is below 1\% of the signal variance, demonstrating that the emulator interpolation error is negligible compared to typical experimental uncertainties. All the associated figures can be found in the GitHub repository~\cite{DominguesUniversalityRepo2025}.
\section{Global Sensitivity Analysis over viscous corrections}
\label{appendix:GSA_viscous_corrections}
In this subsection, we extend the global sensitivity analysis to the other viscous correction schemes considered in this work, as described in Sec.~\ref{subsec:particlization}. While the main text focused on the Grad correction, these complementary analyses serve to validate the robustness of the sensitivity trends and to assess whether the dominant parameters remain consistent across different viscous formulations.

For all cases, the analysis was performed using the same methodology described in Sec.~\ref{subsec:sensitivity_analysis_params}, based on the first principal component (PC1) of the scaled spectra $U(x_T)$. 
This component captures the largest fraction of the variance across the design space and effectively summarizes the global response of the model to parameter variations.
The Sobol first-order indices were calculated using Gaussian process emulators trained on the respective model outputs.

Overall, the CE and PTM viscous corrections exhibit sensitivity patterns qualitatively similar to those observed for the Grad model: the free-streaming time scale $\tau_R$, the maximum bulk viscosity $(\zeta/s)_{\rm max}$, the temperature of ($\zeta/s$) at the peak $T_{\zeta}$, and the nucleon width $w$ remain the most influential parameters controlling the shape of the scaled spectra. 
In contrast, the PTB correction shows a somewhat different behavior, with reduced sensitivity to $\tau_R$ and enhanced dependence on bulk-viscous effects, consistent with its distinct functional form that exponentiates the viscous correction to the distribution function (Eq.~\ref{eq:PTB_equation}).
\begin{figure}[ht]
  \centering
  \includegraphics[width=\linewidth]{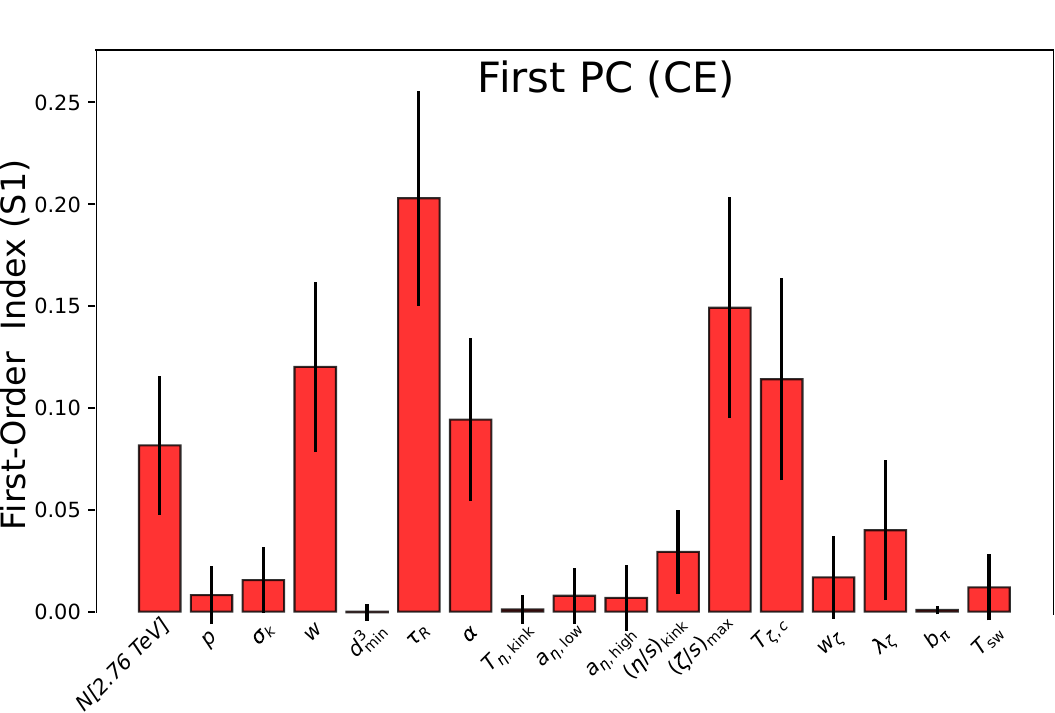}
  \caption{First-order Sobol sensitivity indices for the Chapman–Enskog (CE) viscous correction, computed using the first principal component (PC1) of the scaled spectra $U(x_T)$. The dominant parameters are consistent with those from the Grad correction in Fig.~\ref{fig:Grad_sobol_index}.}
  \label{fig:GSA_CE}
\end{figure}

\begin{figure}[ht]
  \centering
  \includegraphics[width=\linewidth]{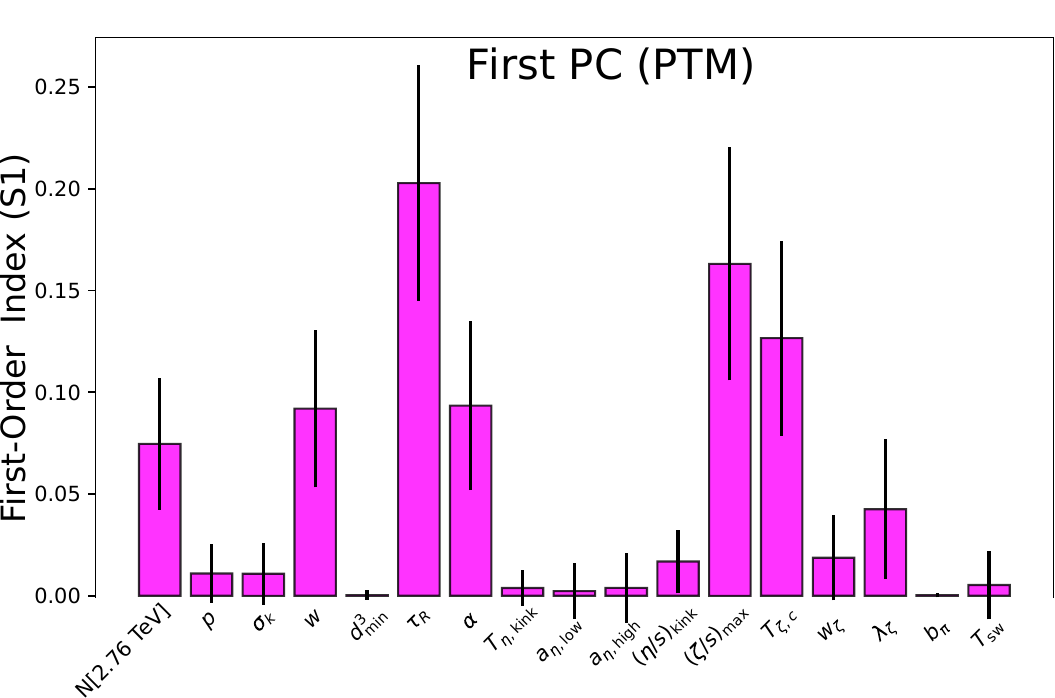}
  \caption{First-order Sobol sensitivity indices for the Pratt–Torrieri–McNelis (PTM) viscous correction. The sensitivity pattern is similar to Grad and CE, with $\tau_R$, $(\zeta/s)_{\rm max}$, and $T_{\zeta}$ as the three most influential parameters. However, $\alpha$ shows a slightly larger sensitivity than $w$ in this case.}
  \label{fig:GSA_PTM}
\end{figure}

\begin{figure}[ht]
  \centering
  \includegraphics[width=\linewidth]{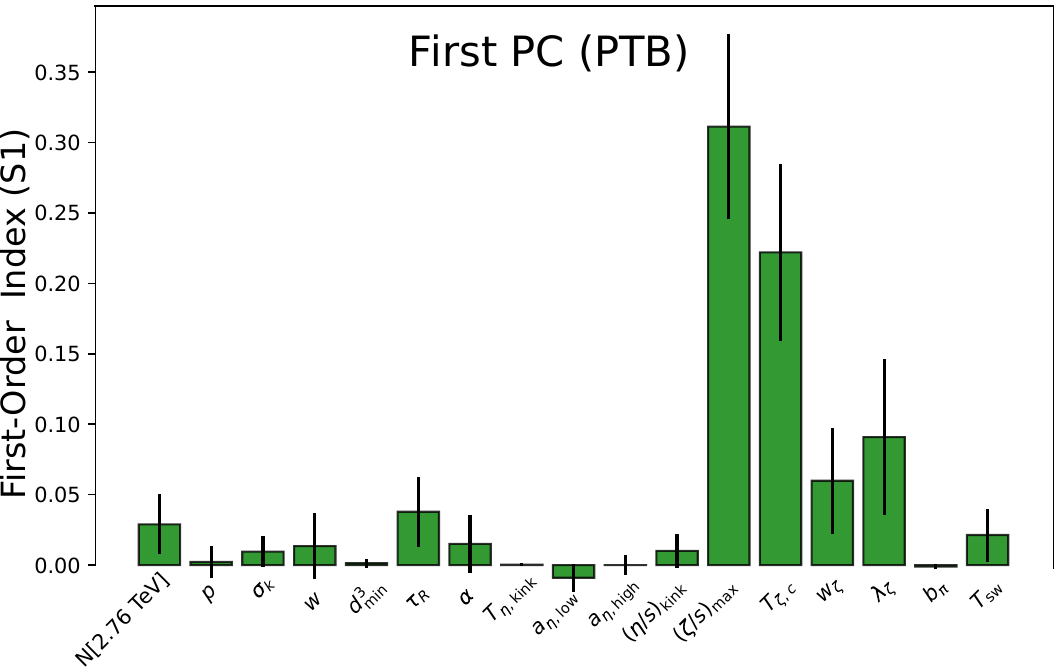}
  \caption{First-order Sobol sensitivity indices for the Pratt–Torrieri–Bernhard (PTB) viscous correction. Compared to the other $\delta f$ models, PTB exhibits stronger dependence on bulk viscosity parameters and weaker sensitivity to $\tau_R$.}
  \label{fig:GSA_PTB}
\end{figure}
These results demonstrate that, despite quantitative differences among viscous corrections, the overall hierarchy of parameter influence remains stable for Grad, CE, and PTM. 
The PTB correction stands out due to its non-linear exponentiation of viscous effects, leading to larger uncertainties and a distinct sensitivity pattern, in line with the emulator validation results discussed in Appendix~\ref{appendix:emulator_PCA}.

\bibliography{ref.bib}
\end{document}